\documentclass[fleqn,usenatbib]{mnras}

\usepackage{newtxtext,newtxmath}

\usepackage[T1]{fontenc}

\DeclareRobustCommand{\VAN}[3]{#2}
\let\VANthebibliography\thebibliography
\def\thebibliography{\DeclareRobustCommand{\VAN}[3]{##3}\VANthebibliography}

\usepackage{graphicx}	
\usepackage{amsmath}	
\usepackage{bm}         
\usepackage{pifont}

\usepackage{xcolor}

\newcommand{\ketju}{\textsc{Ketju}}                           
\newcommand{\gadget}{\textsc{gadget-4}}                       

\newcommand{\Msun}{\ensuremath{\mathrm{M}_{\sun}}}            
\newcommand{\kmps}{\ensuremath{\mathrm{km}\,\mathrm{s}^{-1}} }
\newcommand{\Reff}{\ensuremath{R_\mathrm{e}}}                 
\newcommand{\rb}{\ensuremath{r_\mathrm{b}}}                   
\newcommand{\rbbin}{\ensuremath{r_\mathrm{b,0}}}                   
\newcommand{\vk}{\ensuremath{v_\mathrm{kick}}}                
\newcommand{\vkmax}{\ensuremath{v_\mathrm{kick, max}}}        

\newcommand{\dd}[1]{\ensuremath{\mathrm{d}#1}}                
\newcommand{\dnv}[3]{\ensuremath{\frac{\mathrm{d}^#1#2}{\dd{#3}^#1}}}  
\newcommand{\vb}[1]{\ensuremath{\bm{#1}}}                     
\newcommand{\ordinal}[1]{\ensuremath{#1^\mathrm{th}}}                     
\newcommand{\ordinalthird}[1]{\ensuremath{#1^\mathrm{rd}}}                     

\graphicspath{{figures/}}

\title[Identifying SMBH recoil]{
Identifying supermassive black hole recoil in elliptical galaxies
}

\author[A. Rawlings et al.]{Alexander Rawlings,$^{1}$\thanks{E-mail: alexander.rawlings@helsinki.fi}
Atte Keitaanranta,$^{1}$
Max Mattero,$^{1}$
Sonja Soininen,$^{1}$
Ruby J. Wright,$^{2,1}$
\newauthor
Noa Kallioinen,$^{3}$
Shihong Liao,$^{4}$
Antti Rantala,$^{5}$
Peter H. Johansson,$^{1}$
Thorsten Naab,$^{5}$ and
\newauthor Dimitrios Irodotou$^{6}$
\vspace*{0.1cm}\\
$^{1}$
Department of Physics,
Gustaf H\"allstr\"omin katu 2, FI-00014, University of Helsinki, Finland
\\
$^{2}$
International Centre for Radio Astronomy Research (ICRAR), University of Western Australia, Crawley, WA 6009, Australia
\\
$^{3}$
Department of Computer Science, Konemiehentie 2, 02150, Aalto University, Espoo, Finland
\\
$^{4}$
Key Laboratory for Computational Astrophysics, National Astronomical Observatories, Chinese Academy of Sciences, Beijing 100101, China
\\
$^{5}$
Max-Planck-Institut f\"ur Astrophysik, Karl-Schwarzchild-Str 1, D-85748 Garching, Germany
\\
$^{6}$
The Institute of Cancer Research, 123 Old Brompton Road, London SW7 3RP, United Kingdom
\\
}

\date{Accepted XXX. Received YYY; in original form ZZZ}

\pubyear{2024}

\begin{document}
\label{firstpage}
\pagerange{\pageref{firstpage}--\pageref{lastpage}}
\maketitle

\begin{abstract}
We study stellar core growth in simulations of merging massive ($M_\star>10^{11}\,\Msun$) elliptical galaxies by a supermassive black hole (SMBH) displaced by gravitational wave induced recoil velocity.
With controlled, dense sampling of the SMBH recoil velocity, we find the core radius originally formed by SMBH binary scouring can grow by a factor of 2-3 when the recoil velocity exceeds $\sim50$ per cent of the central escape velocity, and the mass deficit grows by up to a factor of $\sim4$.
Using Bayesian inference we predict the distribution of stellar core sizes formed through this process to peak at $\sim1\,\mathrm{kpc}$.
An orbital decomposition of stellar particles within the core reveals that radial orbits dominate over tube orbits when the recoil velocity exceeds the velocity dispersion of the core, whereas tube orbits dominate for the lowest recoil kicks.
A change in orbital structure is reflected in the anisotropy parameter, with a central tangential bias present only for recoil velocities less than the local stellar velocity dispersion.
Emulating current integral field unit observations of the stellar line-of-sight velocity distribution, we uncover a distinct signature in the Gauss-Hermite symmetric deviation coefficient $h_4$ that uniquely constrains the core size due to binary scouring.
This signature is insensitive to the later evolution of the stellar mass distribution due to SMBH recoil.
Our results provide a novel method to estimate the SMBH recoil magnitude from observations of local elliptical galaxies, and implies these galaxies primarily experienced recoil velocities less than the stellar velocity dispersion of the core.
\end{abstract}




\section{Introduction}
Observations have revealed a clear bimodality in the properties of massive elliptical galaxies, with brighter ellipticals $(M_{B}\lesssim -20.5)$ typically having `cored' flat central surface brightness profiles, boxy isophotes and showing little or no rotation (slow rotators). Fainter intermediate-mass ellipticals $(-20.5 \lesssim M_{B} \lesssim -18.5)$ on the other hand typically show steeper `cuspy' central brightness profiles, more disc-like isophotes and evidence for more rotational support (fast rotators) (e.g. \citealt{Kormendy1996,Faber1997,emsellem2007,Kormendy2009,Thomas2014,dullo2014,dullo2019,Quenneville2024}).

The observed bimodality can be linked to the distinct formation paths of elliptical galaxies. Intermediate-mass ellipticals have likely formed predominantly through in-situ star formation and in mergers of gas-rich disc-dominated galaxies accompanied by merger-induced starbursts resulting in cuspy central stellar profiles (e.g. \citealt{Barnes1996,Cappellari2007,hopkins2009a,hopkins2009b,Johansson2009,Krajnovic2011,Cappellari2016,Lahen2018,Forster2020}). 
More massive elliptical galaxies are on the other hand believed to have assembled in a two-stage process, in which the early assembly at redshifts of $z \gtrsim 1.5$ is dominated by in-situ star formation fuelled by cold gas flows in massive dark matter (DM) haloes, and the accretion of multiple star-bursting progenitors. The later evolution of the galaxies then proceeds mainly through gas-poor (dry) merging in which the galaxy predominantly accretes stars formed  outside the main galaxy (e.g. \citealt{Naab2009,Bezanson2009,Oser2010,Johansson2012,Wellons2015,Rodriguez-Gomez2016,Qu2017,lagos2022,Cannarozzo2023}, see also \citealt{naab2017} for a review). However, if massive galaxies have formed by merging of fainter cuspy galaxies they should also possess central cusps, as simulations have shown that mergers tend to preserve the density cusps of merging galaxies \citep{boylan2004,Dehnen2005}.          

In the case of two massive elliptical galaxies undergoing a dry merger, the leading mechanism for stellar core formation is through the interaction of binary supermassive black holes (SMBHs) in the galaxy merger remnant. The merging of SMBHs is generally understood in the context of a three-stage process \citep{begelman1980}. First, the dynamical friction from the mass overdensity in the wakes of the SMBHs provides a restoring force which brings the SMBHs to the centre of the two interacting galaxies \citep{chandrasekhar1943}, generally torquing the SMBHs to highly radial orbits in the process.
Second, after the mass enclosed within the orbit of the two SMBHs is comparable to the sum of the SMBH masses, the SMBHs form a bound binary \citep{merritt2005}.
Iterative interactions of the SMBHs with surrounding stars on low angular momentum orbits (the stellar `loss-cone' population) eject these stars with high velocity \citep[e.g.][]{hills1980,hills1983,quinlan1996}, thus removing orbital energy and angular momentum from the SMBH binary orbit.
The removal of stars from the central regions by this slingshot mechanism produces a `core' \citep[e.g.][]{lauer1983,lauer1985,kormendy1984}, or depletion, in the luminosity profile of the remnant galaxy \citep[e.g.][]{begelman1980,hills1983,quinlan1996,rantala2018,rantala2024,nasim2021b,khonji2024,partmann2024}, which can in extreme cases extend up to $3\,\mathrm{kpc}$ \citep{postman2012}.
Early numerical simulations found that during the scouring of the stellar core, the loss-cone population of stars can be exhausted if the galactic nucleus has a high degree of spherical symmetry, thus preventing the SMBH binary from reaching separations below roughly one parsec.
This phenomenon is termed the final-parsec problem \citep{milosavljevic2001}.
Subsequent work found that, in the gas-free case, stellar orbit diffusion from more realistic triaxial potentials prevents the loss-cone from emptying, allowing the SMBH binary to reach sub-parsec separations \citep[e.g.][]{berczik2006, holley-bockelmann2006, merritt2011, vasiliev2015, gualandris2017}.
At these very small separations the SMBH binary loses its remaining orbital energy and angular momentum through gravitational wave (GW) emission, driving the SMBH binary to coalescence \citep{peters1963,peters1964}. The GW emission from such SMBH coalescence events are prime observational targets for ongoing pulsar timing array (PTA) detectors for SMBHs with masses of $M_\bullet \gtrsim 10^9\,\Msun$ \citep{agazie2023, antoniadis2024, 2023Xu, zic2023}, and also the forthcoming Laser Interferometer Space Antenna (LISA) mission for SMBHs with slightly lower masses of $M_\bullet \lesssim 10^8\,\Msun$ \citep{amaro-seoane2023}.

The final GW emission from the coalescing SMBH binary occurs anisotropically, and carries with it linear momentum from the system \citep[e.g.][]{gonzalez2007}.
This results in the remnant SMBH recoiling with some velocity directed opposite to the linear momentum of the GW emission \citep{bekenstein1973}, and is termed the kick, or recoil, velocity\footnote{We use the terms `kick' and `recoil' interchangeably in this work.}. 
The kick velocity is dependent on the mass ratio between the SMBHs prior to coalescence, and the magnitude and direction of the angular momentum (spin) vector of each SMBH. 
As reported by \citet{campanelli2007}, symmetries in the masses or spins of the SMBHs suppresses the resulting recoil kick imparted to the coalesced SMBH.
In particular, it is found that asymmetry in the spins of the SMBHs has a larger impact on the magnitude of the recoil kick than asymmetry in the masses \citep{campanelli2007}, with spins which are anti-aligned in the orbital plane and maximal producing the largest recoil kick velocities \citep{gonzalez2007b,tichy2007}.
Numerical relativity studies indicate that while the majority of recoil kicks are of the order of a few hundred kilometres per second, in special configurations the SMBH may be imparted a kick velocity in excess of $2000\,\kmps$, even up to $4000\,\kmps$ \citep{campanelli2007,gonzalez2007b,tichy2007}.
Taking the central escape velocity of a typical massive elliptical galaxy as $\sim2000\,\kmps$, there may be a non-negligible fraction of elliptical galaxies with an SMBH that has escaped the galaxy (e.g. \citealt{madau2004,mannerkoski2022}), though not so many so as to introduce considerable scatter into the observed relation between SMBH mass and stellar velocity dispersion \citep{volonteri2007}.

Numerous observational studies have also hinted at the existence of recoiling SMBHs, generally through either an offset from the centre of the host galaxy in velocity, in position, or both. One of the earliest observations that pointed to a recoiling SMBH was of the quasar SDSSJ0927+2943, reported by \citet{komossa2008}, where an observed velocity offset suggested a recoil kick in excess of $2600\,\kmps$.
Another recoiling SMBH candidate was identified by an offset in velocity by \citet{steinhardt2012}, with an estimated recoil velocity greater than $4000\,\kmps$.
Other observations by \citet{comerford2014} and \citet{pesce2018, pesce2021} have also reported potential recoiling SMBHs, albeit at much lower kick velocities (some tens to few hundreds of $\kmps$).
Recoiling SMBHs may also be identified through a spatial offset from the host centre, where the offset can range from some parsecs \citep{batcheldor2010,lena2014,barrows2016} to over a kiloparsec \citep{koss2014,skipper2018}.
Additionally, some candidate recoiling SMBHs have both spatial and velocity offsets, such as CXO J101527.2+625911 \citep{kim2017} and 2MASX J00423991+3017515 \citep{hogg2021}.
Attributing an escaping SMBH to GW recoil as opposed to three body interactions between an SMBH binary and a third SMBH is also not trivial, as demonstrated in the recent observations presented in \citet{vandokkum2023}.

Numerically, the effect of recoiling SMBHs on the formation of cores in massive elliptical galaxies has been 
primarily studied using collisionless simulations. Early work by \citet{boylan2004}
found that a recoiling SMBH induced a core in the stellar density profile, particularly when the SMBH remained bound to the bulge of the host galaxy and had a recoil velocity comparable to the bulge velocity dispersion. In another early study 
\citet{merritt2004} also found that recoiling SMBHs facilitated the formation of cores
and found that ejected SMBHs settled roughly in one orbital period in spherical potentials, with this `infall time' increasing when more realistic triaxial potentials were considered. 
In \citet{gualandris2008} the motion of a settling SMBH in a stellar potential representative of a galaxy merger remnant was found to occur in three stages: first, the SMBH oscillates with decreasing amplitude due to energy dissipation by dynamical friction.
Next, both the SMBH and stellar core oscillate about their common centre of mass when the radial excursion of the SMBH is less than the core radius of the stellar system.
Finally, the SMBH settles into thermal equilibrium with the surrounding stars.
During these three stages, the stellar density core is enhanced by the motion of the SMBH, producing a mass deficit in some cases up to five times the SMBH mass (see also \citealt{merritt2004,boylan2004,partmann2024}).
However, a recent study by \citet{nasim2021b} on the contribution to stellar core expansion through GW recoil of an SMBH found intriguingly that the greatest contribution to the stellar mass deficit is due to the first excursion of the SMBH following its ejection, with subsequent oscillations only providing a minor contribution to the growth of the stellar core. 

Whilst previous studies have convincingly shown that there is a clear case for a relation between SMBH recoil velocity and the final core size carved out by its motion, little attention has been devoted to understanding the statistical impact of SMBH recoil on the properties of the merger remnant galaxy, for example the distribution of core sizes that result from realistic SMBH recoil velocities.
Work by \citet{nasim2021b} and \citet{khonji2024} considered specific examples of local massive ETGs, invoking large SMBH recoil velocities with up to 90 per cent of the galaxy escape velocity as a mechanism to grow the most massive observed cores.
These works also considered varying SMBH masses and stellar density profiles, affecting the contribution to core growth by SMBH binary scouring.
Whilst successful in explaining extreme core sizes, such high recoil velocities are unlikely to be the norm, given that all observed elliptical galaxies host a central SMBH \citep{ferrarese1994,haring2004,kormendy2013,vandenbosch2016}.
The role of more typical SMBH recoil velocities \citep[of the order a few to several hundred kilometres per second, e.g.][]{campanelli2007} in core formation in general, and not just in the case of the most massive cores, remains largely unexplored.
In addition, how a recoiling SMBH impacts the surrounding stellar orbital distribution has remained largely unanswered. 
Orbital modelling techniques, in particular orbit-superposition Schwarzschild modelling, offer a unique bridging between observations and simulations, often revealing a rich understanding of a galaxy's history, and the effect of SMBH interactions \citep[e.g.][]{rottgers2014,neureiter2021,neureiter2023,santucci2022,santucci2023,santucci2024}.

In this study we use the \gadget{} based regularised tree code \ketju{} \citep{rantala2017,rantala2020,mannerkoski2023} to study the effect of recoiling SMBHs on the structure of massive elliptical galaxies. 
Previously, \ketju{} has been used to study the formation of cores through SMBH binary scouring in gas-free simulations, but without the contribution of SMBH recoils \citep{rantala2018,rantala2019,rantala2024}.
Here, we run a large set of gas-free simulations with systematically varying SMBH kick velocities.  
The aim of this work is twofold.
We first aim to determine how the velocity with which the SMBH is kicked affects the mass distribution of the surrounding stellar environment, and the likelihood of the magnitude of these effects. We then aim to determine if a recoiling SMBH imparts a signature on the stellar kinematics of the remnant galaxy, and if such a signature could potentially be observed with current detectors.

This paper is divided as follows: we present the numerical simulations used in our investigation in \autoref{sec:num_sims}.
We discuss the motion of the recoiling SMBHs in \autoref{sec:motion}, and the resulting stellar density profiles of the remnant galaxy in \autoref{sec:density}.
We investigate the distribution of stellar orbits as a function of the kick velocity in \autoref{sec:orbits} and construct mock stellar kinematic observations to be compared with observational data in \autoref{sec:kinematics}.
Finally, we discuss our results in \autoref{sec:discussion} and present 
our conclusions in \autoref{sec:conclusions}, respectively.

\section{Numerical Simulations}
\label{sec:num_sims}

\subsection{Simulation Code}
To investigate the effect of the recoiling SMBH on the host galaxy, and potential observational signatures of this, we run a number of numerical simulations of an idealised galaxy merger setting using the new public version of the \ketju{}\footnote{\url{https://www.mv.helsinki.fi/home/phjohans/group-website/research/ketju/}} code \citep{rantala2017,rantala2020,mannerkoski2023} coupled with \gadget{} \citep{springel2021}.
\ketju{} integrates the dynamics of SMBHs, and stellar particles in a small region three times the stellar softening length around them (i.e. the \ketju{} region), to high accuracy using the algorithmically regularised integrator \textsc{mstar} \citep{rantala2020}.
The dynamics of stellar particles beyond this small region of high integration accuracy, and all DM particles, is computed with the \gadget{} fast multiple method (FMM) with second order multipoles. 
Additionally, we use hierarchical time integration, which allows for mutually symmetric interactions and manifest momentum conservation. 
In this study, we do not use comoving integration, as we are investigating an idealised system.
\ketju{} also includes post-Newtonian (PN) correction terms up to order 3.5 between each pair of SMBHs \citep{blanchet2014} and the fitting formula of \citet{zlochower2015} for recoil kick velocity, making \ketju{} a particularly well-suited code for investigating the consequence of SMBH recoil self-consistently in a galaxy merger environment. 

\subsection{Initial Conditions}
We model the merger of two massive elliptical galaxies using gas-free merger simulations.
Our galaxy initial conditions follow that of IC $\gamma$-1.0-BH-6 presented in \citet{rantala2018} but scaled down by a factor of $\sim 3$ in mass, consistent with the method presented in \citet{rantala2019}.
As we are interested in the evolution of the galaxy merger \textit{remnant} following SMBH coalescence, the initial conditions are constructed such that at the time of SMBH coalescence (\autoref{ssec:mergers}) the galaxy remnant agrees with local scaling relations.
Specifically, we ensure that our remnant agrees with the half-light -- stellar mass data presented in \citet{sahu2020}, and lies within the $1\sigma$ prediction interval of the SMBH mass -- stellar velocity dispersion data as given in \citet{vandenbosch2016}.

The galaxy is represented as an isotropic multicomponent sphere, and consists of a stellar component of total mass $M_\star\sim 1.38\times10^{11}\,\Msun$ embedded within a DM component of total mass $M_\mathrm{DM}=2.5\times10^{13}\,\Msun$.
At the centre an SMBH with mass $M_{\bullet,0}=2.93\times10^{9}\,\Msun$ is placed with zero velocity.
The stellar and DM components each follow a \citet{hernquist1990} profile with a scale radius of $a_\star=3.9\,\mathrm{kpc}$ and $a_\mathrm{DM}=245\,\mathrm{kpc}$, respectively.
The density profile $\rho_i$ for a given component $i$ is given by:
\begin{equation}\label{eq:dehnen}
    \rho_i(r) = \frac{M_i}{2\pi} \frac{a_i}{r (r+a_i)^{3}}.
\end{equation}
We generate the ICs using the distribution function method following \citet{hilz2012} and \citet{rantala2017}, where for each component (stellar and DM) the distribution function $f_i$ is computed using Eddington's formula \citep{binney2008} for each density profile $\rho_i$:
\begin{equation}
    f_i(\mathcal{E}) = \frac{1}{2\sqrt{2}\pi^2} \int_{\Phi_\mathrm{T}=0}^{\Phi_\mathrm{T}=\mathcal{E}} \dnv{2}{\rho_i}{\Phi_\mathrm{T}} \frac{\dd{\Phi_\mathrm{T}}}{\sqrt{\mathcal{E}-\Phi_\mathrm{T}}}.
\end{equation}
Here $\mathcal{E}$ is the relative energy, and $\Phi_\mathrm{T}$ is the total gravitational potential. 
The distribution functions are then sampled with discrete particles, where the mass of a stellar particle is set to $m_\star=5\times10^4\,\Msun$, and the mass of a DM particle is set to $m_\mathrm{DM}=5\times10^6\,\Msun$.
The radial velocity profiles of both stellar and DM particles are ergodic. 

\subsection{Merger Simulations}\label{ssec:mergers}
We merge two independently Monte Carlo sampled galaxy ICs, creating an equal-mass galaxy merger with a near-radial orbit. 
The merger orbit has an initial separation consistent with \citet{rantala2018} of $D=30\,\mathrm{kpc}$, however the initial eccentricity is set to $e_0=0.97$, and the first pericentre distance adjusted accordingly to $r_\mathrm{peri}=D(1-e_0^2) \simeq 2\,\mathrm{kpc}$ \citep{khochfar2006,rantala2017}.
The merger configuration results in a rapid coalescence of the SMBH binary, however we note that the merger timescale we observe is driven by stochasticity in the binary eccentricity \citep{nasim2020,rawlings2023}.
The resulting coalesced SMBH has a mass $M_\bullet = 2M_{\bullet,0} = 5.86\times10^9\,\Msun$.
We then select the snapshot just prior ($\sim8\,\mathrm{Myr}$) to the GW-driven SMBH merger, and generate 31 `child' simulations, where each child has a unique gravitational recoil kick velocity $\vk$ prescribed along the $x$-axis\footnote{As the $x$-axis is in the global coordinate frame, the direction of the kick is essentially random with respect to the angular momentum and inertia tensors. To test the effect of differing kick directions on the merger remnant, we test directing the recoil kick along the global $y$-axis, and find very similar evolution to the $x$-axis case.}, ranging from $0\,\kmps$ to $1800\,\kmps$ (inclusive), in $60\,\kmps$ increments.
The upper limit on $\vk$ is chosen to match the escape velocity of the centre of the merger remnant, $v_\mathrm{esc}=1800\,\kmps$.
We additionally run one simulation with a recoil kick above $v_\mathrm{esc}$, with $\vk=2000\,\kmps$, to test the effect of rapidly removing an SMBH.

For all simulations, snapshots are produced every $5\,\mathrm{Myr}$.
Interactions between stellar particles are softened with a softening length of $\varepsilon_\star = 2.5\,\mathrm{pc}$ following \citet{rawlings2023}, DM is softened with a softening length of $\varepsilon_\mathrm{DM} = 300\,\mathrm{pc}$, and the \ketju{} region radius is set to $r_{\rm ketju}=3\varepsilon_\star = 7.5\,\mathrm{pc}$ to ensure that the stellar-SMBH interactions are always non-softened \citep{rantala2017}.
Within the \ketju{} region, dynamical interactions between stellar particles and SMBHs, and SMBHs-SMBHs are unsoftened.
For the \gadget{} component of the simulation, the integration error tolerance is set to 0.002, whereas the Gragg-Bulirsch-Stoer (GBS) tolerance for \ketju{} is set to $10^{-8}$.

We impose a maximum integration time on all simulations, corresponding to three dynamical times $t_\mathrm{dyn}$, defined in terms of the virial radius $r_{200}$ and virial velocity $v_{200}$ of the galaxy merger remnant as:
\begin{equation}
    t_\mathrm{dyn} = \frac{r_{200}}{v_{200}} \simeq 1\,\mathrm{Gyr}.
\end{equation}
Thus, our simulations are run for a maximum of $3\,\mathrm{Gyr}$.

\section{Settling SMBHs}\label{sec:motion}

\subsection{Motion of the kicked SMBH}

\begin{figure}
    \centering
    \includegraphics[width=0.48\textwidth]{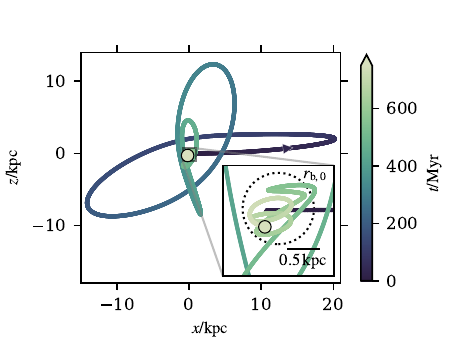}
    \caption{
    The trajectory of the kicked SMBH in the $\vk=780\,\kmps$ simulation, with its final position shown by the marked circle.
    The line is coloured by the time following merger, and an arrow displays the initial outward radial motion of the SMBH for clarity.
    The inset panel depicts the motion of the SMBH as it is settling within the binary-scoured core radius, marked by the dotted circle.
    The final settling of the SMBH is chaotic once its velocity falls below the local stellar velocity dispersion: it is unable to leave the core again from this moment onwards.
    }
    \label{fig:trajectory}
\end{figure}

Before the SMBH binary coalescence, the binary scours a stellar core with a radius of $\rbbin\sim0.58\,\mathrm{kpc}$ in the centre of the galaxy (see \autoref{ssec:projdens}). In the interior of the core, the stellar one-dimensional velocity dispersion is $\sigma_{\star,0} \equiv \sigma_\star(r<\rbbin)\sim 270\,\kmps$. 
Immediately following the coalescence of the SMBH binary and the subsequent kick, the trajectory of the remnant SMBH is predominantly radial in the direction of the GW recoil kick, as shown in \autoref{fig:trajectory}.
We can thus estimate the crossing time of the binary-scoured core based on the 1D stellar velocity dispersion as:
\begin{equation}
    t_\mathrm{cross, core} \simeq \frac{\rbbin}{\sigma_{\star,0}} \approx 2\,\mathrm{Myr},
\end{equation}
whereas the actual time the recoiling SMBH traverses the core is:
\begin{equation}
    t_\mathrm{cross, \bullet} \simeq \frac{\rbbin}{\vk}.
\end{equation}
If the velocity of the recoil kick is not high enough to completely eject the SMBH from the galaxy ($\vk < v_\mathrm{esc}$), the SMBH will oscillate about the centre of the galaxy. 
The amplitude of the oscillations, termed the apocentre distance $r_\mathrm{apo}$, decreases with each oscillation due to dynamical friction, which is automatically resolved with \ketju{} \citep[e.g.][]{karl2015,rantala2017,genina2024}.
Ultimately the SMBH `settles' to the centre of the galaxy.
If $\vk < \sigma_{\star,0}$, the SMBH has insufficient kinetic energy to escape the stellar core and instead remains within $100\,\mathrm{pc}$ from the centre, shown in the top panel of \autoref{fig:infall}.
Conversely, if $\vk > \sigma_{\star,0}$ the SMBH can exit the core, with $r_\mathrm{apo}$ increasing as $\vk$ is increased.
If $r_\mathrm{apo}$ exceeds $\sim 10\,\mathrm{kpc}$, mild triaxiality in the outer regions of the galaxy merger remnant (shown in \autoref{fig:triax}) can torque the SMBH into an orbit with some degree of tangential motion, an example of which is given in \autoref{fig:trajectory} for the $\vk=780\,\kmps$ simulation.
For the first few orbits, the SMBH may thus not necessarily pass through the stellar core, though it loses energy to the surrounding medium with each oscillation.
During the final passages however, the SMBH passes through the core, before its velocity decreases below the velocity dispersion of the core.
The predominantly-radial motion of the SMBH rapidly degrades to a random walk as it settles to the Brownian limit from this point, and does not leave the core again (shown in the inset panel of \autoref{fig:trajectory}).

\subsection{Identifying a settled SMBH}\label{ssec:settle}
As we focus on galaxy merger remnants and the possible observational signatures caused by recoil kicks in this work, we analyse herein times after the SMBH has settled back to the centre of the galaxy. 
Therefore we first need to determine the time (the infall time henceforth) where this occurs. 
The stellar centre of the merger remnant is found using the shrinking sphere method \citep{power2003}: we use this method herein for our analysis.
We begin by estimating the expected Brownian velocity of an SMBH in a distribution of stellar and DM particles of lower unit mass ($m_\star, m_\mathrm{DM} \ll M_\bullet$) following \citet{merritt2007}.
We determine the root mean square velocity of the massive particle as
\begin{equation}
    V_\mathrm{RMS} = \sqrt{3 \xi} \tilde{\sigma},
\end{equation}
where $\tilde{\sigma}$ is the average velocity dispersion of the stellar and DM particles, and $\xi$ is the mass ratio to the SMBH, calculated as:
\begin{equation}
    \xi = \frac{\int n(m)m^2\dd{m}}{\int n(m)m\dd{m}} \Big / M_\bullet,
\end{equation}
where $n(m)$ is the number density of particles with masses $m$ to $m+\dd{m}$.
Our calculated value of $V_\mathrm{RMS}$ that we use to define a settled velocity is $V_\mathrm{RMS} \sim 25\,\kmps$.

Next, we go through the simulations output in `windows' of $100\,\mathrm{Myr}$. 
If the SMBH stays within $0.1\,\mathrm{kpc}$ from the centre and has a median velocity less than $V_\mathrm{RMS}$ relative to the centre within the $100\,\mathrm{Myr}$ window, we consider the SMBH to be settled at the time when the velocity first falls below $V_\mathrm{RMS}$ within the time interval. The chosen separation from the centre is smaller than the large offsets observed by \citet{barrows2016} and slightly larger than the $\lesssim 10\,\mathrm{pc}$ offsets in \citet{lena2014}. 

\begin{figure}
    \centering
    \includegraphics[width=0.4\textwidth]{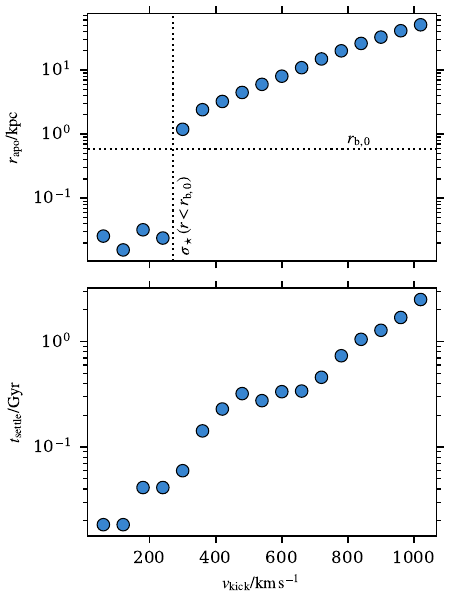}
    \caption{ \textit{Top panel}: the distance of the first apocentre passage from the centre of the galaxy for each kick velocity that resulted in a settled SMBH. 
    The horizontal dashed line, $\rbbin$, marks the size of the stellar core the SMBH binary scours before coalescence, and the vertical dashed line, $\sigma_\star(r<\rbbin)$, is the one-dimensional velocity dispersion of the stellar particles interior to the binary-scoured core radius. \textit{Bottom panel}: the settling times for each kick velocity with $\vk>0\,\kmps$. The time increases exponentially as  $\vk$ increases.
    }
    \label{fig:infall}
\end{figure}

The settling times are shown in the lower panel of \autoref{fig:infall}. For recoil velocities $\vk \leq 240\,\kmps$ the kick is not large enough to displace the SMBH beyond the binary-scoured core and the SMBH is considered settled in less than $100\,\mathrm{Myr}$ after SMBH binary coalescence. 
As the kick velocity is increased further, the settling time begins to increase exponentially. 
The slight deviations from an exponential increase are caused by numerical noise, which prevents the median velocity being below $V_\mathrm{RMS}$.

The SMBHs which received a kick velocity larger than $1020\,\kmps$ did not settle back to the centre of the galaxy within the maximum simulation time of $3t_\mathrm{dyn} = 3\,\mathrm{Gyr}$, and already reach maximum displacements in excess of $>60\,\mathrm{kpc}$ during their first excursion from the centre. 
We expect that any SMBH with such a large displacement from its host galaxy would not be observationally identified with that galaxy, and thus we do not include simulations with $\vk>1020\,\kmps$ in this analysis. The used limit for maximum displacement is a few times larger than the largest suggested offset of $19.4\,\mathrm{kpc}$ from observations \citep{barrows2016}.
We do however analyse the $\vk=2000\,\kmps$ simulation, which is above the escape velocity of the galaxy, as a proxy for an SMBH that is almost instantaneously removed from the simulation.
Importantly, an escaping SMBH never completes a single oscillation about the centre of the galaxy.

\subsection{Movement of stellar mass by a settling SMBH}\label{ssec:movement}
We consider the amount of stellar mass that is bound to the recoiling SMBH, $M_{\star,\mathrm{bound}}$, at different times following coalescence.
A stellar particle is defined to be bound to the SMBH if it is within the influence radius $r_\mathrm{infl}$ of the SMBH, which satisfies \citep[e.g.][]{merritt2013}:
\begin{equation}
    M_\star(r < r_\mathrm{infl}) = 2M_\bullet,
\end{equation}
and the binding energy of the stellar particle $K-U$, where $K$ is the kinetic energy and $U$ is the potential energy (in the SMBH frame of reference), is less than zero.
We determine the SMBH influence radius to be $r_\mathrm{infl}\simeq 1.5\,\mathrm{kpc}$ at the time of SMBH coalescence.
At each pericentre passage of the oscillating SMBH (where the SMBH velocity is maximal), $M_{\star,\mathrm{bound}}$ is minimal; conversely, $M_{\star,\mathrm{bound}}$ is maximal at apocentre, irrespective of $\vk$.
Increasing $\vk$ however reduces the magnitude of $M_{\star,\mathrm{bound}}$ compared to lower values of $\vk$.
Following coalescence for the $\vk=60\,\kmps$ case (the lowest, non-zero recoil velocity simulated), $M_{\star,\mathrm{bound}} \sim 2.18\times10^9\,\Msun$, or roughly 37 per cent of the coalesced SMBH mass and less than 2 per cent of the total stellar mass.
By $\vk=1020\,\kmps$, $M_{\star,\mathrm{bound}}$ has decreased to 0.2 per cent of the coalesced SMBH mass.
These results are in close quantitative agreement with the findings of \citet{gualandris2008} and agree well with those of \citet{nasim2021b}, despite different initial conditions being used between our study and those in the literature.
We hypothesise that the movement of stellar mass bound to the SMBH is not the dominant mechanism by which the stellar mass distribution changes, and investigate this further in \autoref{ssec:missmass}.

\begin{figure}
    \centering
    \includegraphics[width=0.5\textwidth]{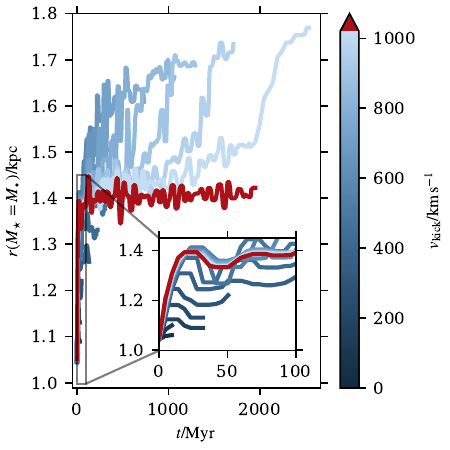}
    \caption{ 
    The radius from the centre of the galaxy which encloses a stellar mass equivalent to the mass of the SMBH as a function of time for each kick velocity. 
    Recoil velocities from $0\,\kmps$ to $1020\,\kmps$ are shown with a continuous colour scale, and the $2000\,\kmps$ simulation is shown in red. For each recoil velocity, the Lagrangian radius is shown until the SMBH has settled. The evolution is shown until $t\sim2.0\,\mathrm{Gyr}$ for the SMBH ejected from the galaxy with a $2000\, \kmps$ kick. 
    During the first $\sim 30\,\mathrm{Myr}$, the Lagrangian radius increases with larger kick velocities resulting in larger growth of the core. For the kick velocities that manage to generate multiple oscillations from the centre, the Lagrangian radius increases further when the SMBH travels through the core region during its final few orbits.
    }
    \label{fig:lagrangian}
\end{figure}

The core region of the galaxy constantly evolves while the SMBH moves through the galaxy.
We inspect how the Lagrangian radius relative to the centre of mass and enclosing a stellar mass equal to the SMBH mass, $r(M_\star = M_\bullet)$, evolves with time in \autoref{fig:lagrangian} until the SMBH is settled. 
In addition to kick velocities $\vk \leq1020\,\kmps$, the kick velocity $\vk=2000\,\kmps$ is included. 
As the SMBH with $\vk=2000\,\kmps$ is completely removed from the galaxy and does not settle back to the centre, the evolution is shown for $\sim 2.0\,\mathrm{Gyr}$ following SMBH binary coalescence. 

During the first $\sim 30\,\mathrm{Myr}$, the initial evolution of the Lagrangian radii is very similar for kick velocities $\vk> 300\,\kmps$.
We can understand this from the core-crossing time argument, where if
\begin{equation}
    \frac{\vk}{\sigma_{\star,0}} > 1 \implies \frac{t_\mathrm{cross, \bullet}}{t_\mathrm{cross, core}} < 1.
\end{equation}
Consequently, for $\vk>\sigma_{\star,0}$, the SMBH is moving faster than the stellar core is able to react to the rapid change in potential caused by the recoiling SMBH.
For these simulations, the Lagrangian radii initially increases sharply to $\sim1.4\,\mathrm{kpc}$ and the SMBH leaves the core almost instantly. 
This large rise occurs during the first oscillation. 
The trend which the Lagrangian radius follows at this phase is caused by the reaction of the core region to the changed gravitational potential \citep{boylan2004}. Each simulation shifts away from the trend when the SMBH returns to the centre. 
For $\vk<\sigma_{\star,0}$, the SMBH settles back to within 100 pc from the centre before the Lagrangian radius can reach $\sim1.4\,\mathrm{kpc}$. 

The higher recoil kick velocities in \autoref{fig:lagrangian} show that the Lagrangian radii can further increase at later times in addition to the initial increase, as long as the SMBH is kicked beyond the core. 
The increase is larger for higher kick velocities, with the Lagrangian radius in the simulation with $\vk=1020\,\kmps$ reaching nearly $1.8\,\mathrm{kpc}$. As was seen from \autoref{fig:trajectory}, the SMBH can miss the core region entirely during the first few orbits. 
The Lagrangian radius is only affected when the SMBH passes through the core, or very near to it. 
For each of the simulations with $\sigma_{\star,0} < \vk \leq 1020\,\kmps$, the final few orbits will always include a passage through the core region. The SMBHs which received larger kick velocities possess more kinetic energy compared to those SMBHs which received smaller kick velocities and will cross the core region more times. Since the SMBH struggles to leave the core region during the final oscillations, the kinetic energy heats up the stellar particles within the central region, causing the additional growth of the Lagrangian radius.

Importantly, the Lagrangian radius for each $\vk$ stays largely constant for times after $t_\mathrm{settle}$. Thus, the core region of the galaxy has had enough time to react to the effect the kicked SMBH has caused at the moment when the SMBH settles at the centre. Based on this, the state of the galaxy merger remnant  at $t=t_\mathrm{settle}$ captures the full effect which the kicked SMBH has on the galaxy.

\subsection{Relaxation time}\label{ssec:relax}
As already discussed, the time required for the kicked SMBH to return and settle at the centre of the merger remnant in general increases with increasing recoil velocity.
To ensure that our results are robust to the variable ending time of the simulations, and are not a manifestation of relaxation processes, we determine the relaxation time of the $\vk=0\,\kmps$ system at different radii.
We conservatively estimate the relaxation time $t_\mathrm{relax}$ at a radius $r$ as \citep{binney2008}:
\begin{equation}\label{eq:relax}
    t_\mathrm{relax} \simeq 2.1 \frac{\sigma r^2}{G \bar{m} \ln \Lambda},
\end{equation}
where $\sigma$ is the particle velocity dispersion within $r$, $\bar{m}$ is the mean particle mass (hence the mean of the stellar and DM particle masses), and $\Lambda$ is the argument of the Coulomb logarithm, given by:
\begin{equation}
    \Lambda = \frac{r \langle v^2 \rangle}{2 G \bar{m}},
\end{equation}
where $\langle v^2 \rangle$ is the mean squared particle velocity within $r$. 
The relaxation time increases quadratically with increasing $r$, and we find for a radius of $r=1.0\,\mathrm{kpc}$ the relaxation time to be $t_\mathrm{relax} \sim 5.0 \, \mathrm{Gyr}$; this increases to $\sim 100\,\mathrm{Gyr}$ by $r=5\,\mathrm{kpc}$. 
Of the simulations where the SMBH was determined to have settled, the maximum time for this to occur was $\sim 2.5\,\mathrm{Gyr}$, well below the relaxation time of the system. 
We  note that \autoref{eq:relax} does not take into account gravitational softening, and thus provides a conservative lower bound for the relaxation time.
This is critical as the \ketju{} region does not include gravitational softening between stellar particles and the SMBH, so it is not a fully-softened simulation.
In practice, we expect the relaxation time to be even longer than reported due to the inclusion of gravitational softening in the simulation for particles outside of the \ketju{} region.
Thus, for the analysis herein, we are confident that the results we observe are directly caused by the influence of the recoiling SMBH on the galaxy, and are not due to relaxation effects.

\section{Stellar mass density profiles}\label{sec:density}
\subsection{Three dimensional mass density profiles}

\begin{figure}
    \centering
    \includegraphics[width=0.5\textwidth]{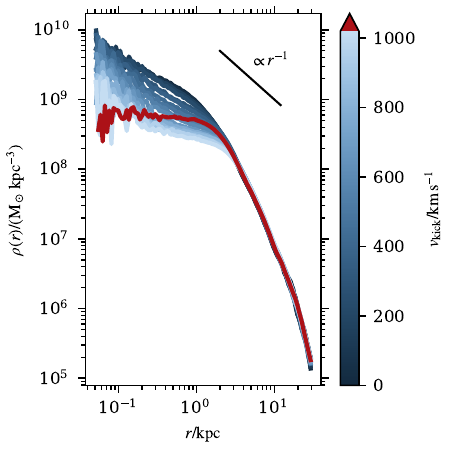}
    \caption{
        Three dimensional stellar mass density, with the line colour corresponding to the kick velocity of the SMBH. 
        At radii $r\lesssim3\,\mathrm{kpc}$, the stellar mass density decreases smoothly with increasing kick velocity, with the exception of the $\vk=2000\,\kmps$ simulation, in which the SMBH has not returned to the centre of the merger remnant.
        A density cusp is apparent for all mergers where the SMBH settled at the centre.
        At radii $r>2\,\mathrm{kpc}$ all the density profiles are consistent with each other, indicating that a changing SMBH recoil velocity has a larger impact on the inner mass distribution of the galaxy merger remnant.
        }
    \label{fig:density3d}
\end{figure}

Before considering the projected, 2D stellar mass density profiles of the merger remnants, we investigate the 3D stellar mass distribution.
This alleviates the influence of projection effects on inferring if a core has formed during the SMBH binary merger process, and gives insight into the dependence of how matter is spatially distributed as a function of the recoiling SMBH kick velocity.

In \autoref{fig:density3d}, we observe a clear decrease in stellar mass density at radii $r\lesssim2\,\mathrm{kpc}$ with increasing kick velocity.
For $\vk=0\,\kmps$, the central stellar density is $\sim10^{10}\,\Msun\,\mathrm{kpc}^{-3}$, whereas for $\vk=1020\,\kmps$ it is $\sim10^{9}\,\Msun\,\mathrm{kpc}^{-3}$.
An exception to this trend is the highest kick velocity we simulated, $\vk=2000\,\kmps$, which shows an excess of stellar mass between $0.3 \lesssim r/\mathrm{kpc} \lesssim 2$ relative to the simulations with $\vk \gtrsim 300\,\kmps$.
This is readily explained by the evolution of the Lagrangian radii in \autoref{fig:lagrangian}.
The final few orbits of the settling SMBH in simulations with $\vk>300\,\kmps$ take the SMBH through the stellar core, enlarging it further with each oscillation. 
As a result, the central stellar density decreases as the number of oscillations increase. 
The $\vk=2000\,\kmps$ simulation, in which the SMBH does not oscillate through the core, has a central stellar density that decreases only due to the initial response of the stellar particles to the sudden change of the potential.
The simulations with a settled SMBH have a density profile within $r<1\,\mathrm{kpc}$ that follows a power law $\rho \propto r^{-\nu}$ with $0<\nu<1$.
Conversely, the $\vk=2000\,\kmps$ simulation has a flat density profile with $\nu=0$ in the centre, in agreement with the recent work by \citet{khonji2024}.

In the outer regions $r\gtrsim 4\,\mathrm{kpc}$, the stellar density falls with increasing radius consistently between all simulations, indicating that the dynamical effect of a recoiling SMBH on the stellar mass distribution is predominantly confined to the central regions of the galaxy.

\subsection{Projected mass density profiles}\label{ssec:projdens}

\begin{figure}
    \centering
    \includegraphics[width=0.4\textwidth]{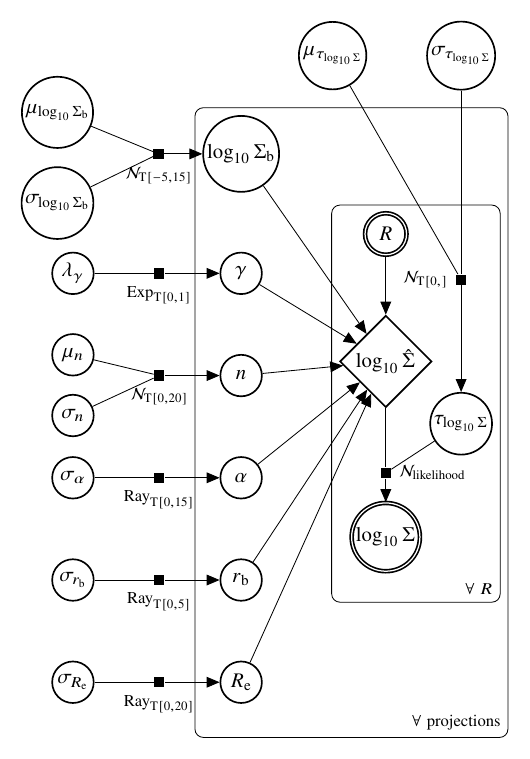}
    \caption{Directed acyclic graph of the core-S\'ersic model of projected mass density within the Bayesian hierarchical framework. Single-line circle nodes represent fit parameters, double-line circles represent measured quantities (the data), and diamond nodes represent deterministic quantities. The particular distribution connecting nodes is written below the corresponding black square, with a subscript `$\mathrm{T}[l,u]$' denoting a distribution $f(\lambda)$ truncated to $l<\lambda<u$, and the subscript `likelihood' indicating the likelihood function. The $R$ box indicates variables fit for each radial bin, and the projection box distinguishes variables specific to each projection realisation. Note that we additionally use $\hat{\Sigma}$ to distinguish the calculated value of surface density from the measured value $\Sigma$. The various distributions are normal ($\mathcal{N}$), exponential ($\mathrm{Exp}$), and Rayleigh ($\mathrm{Ray}$).}
    \label{fig:dag}
\end{figure}

\begin{figure}
    \centering
    \includegraphics[width=0.4\textwidth]{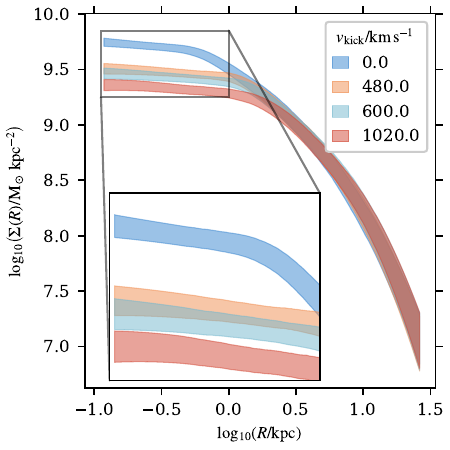}
    \caption{
        Surface density profiles with 25 per cent Bayesian HDI for select representative runs.
        Increasing the kick velocity induces a shallower density profile in the inner regions, consistent with the three dimensional density maps in \autoref{fig:density3d}.
        }
    \label{fig:density}
\end{figure}

\begin{figure}
    \centering
    \includegraphics[width=0.4\textwidth]{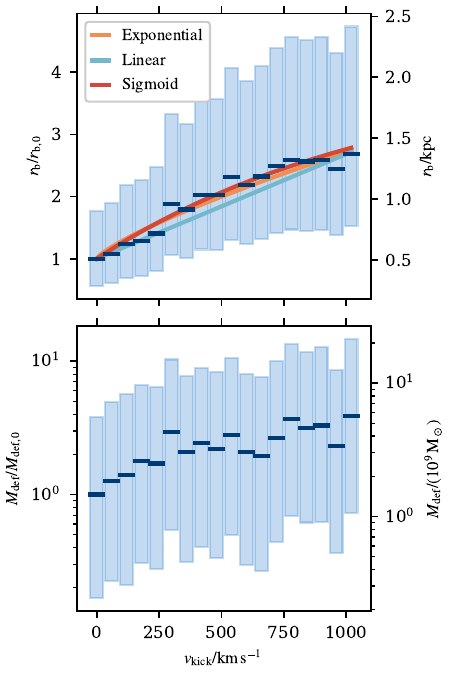}
    \caption{
        \textit{Top panel}: Bayesian estimate of the merger remnant core size $r_\mathrm{b}$, scaled to the core size scoured solely by the SMBH binary prior to coalescence $\rbbin$ (left axis) and in physical units (right axis), as a function of kick velocity $\vk$.
        The core size distributions are shown as box plots, with the median core size indicated by the central mark and shaded regions corresponding to the IQR.
        Larger kick velocities are correlated with larger core sizes. Additionally, a greater spread in the distribution of core sizes over different viewing projections of the merger remnant is associated with larger kick velocities.
        Three models, fit to the median $\rb$--$\vk$ trend, are overplotted.
        For the models, we do not show the parameter uncertainty for visual clarity.
        \textit{Bottom panel}: Box plots of the Monte Carlo sampled mass deficit $M_\mathrm{def}$ between S\'ersic and core-S\'ersic fits as a function of the recoil velocity. 
        $M_\mathrm{def}$ is scaled to the mass deficit due to SMBH binary scouring ($M_{\mathrm{def},0}$, left axis) and shown in physical units (right axis). 
        }
    \label{fig:vkrb}
\end{figure}

To facilitate comparisons with observations \citep[e.g.][]{graham2003,rusli2013}, we fit the six-parameter core-S\'ersic profile \citep{graham2003} to the projected stellar mass density of each simulated merger remnant:
\begin{equation}\label{eq:cs}
    \Sigma(R) = \Sigma' \left[1 + \left(\frac{\rb}{R}\right)^\alpha \right]^{\gamma/\alpha} \exp\left[-b_n \left(\frac{R^\alpha + \rb^\alpha}{\Reff^\alpha}\right)^{(1/\alpha n)}\right]
\end{equation}
where
\begin{equation}\label{eq:cs_prefactor}
    \Sigma' = \Sigma_\mathrm{b} 2^{-\gamma/\alpha} \exp\left[b_n \left(2^{1/\alpha} \frac{\rb}{\Reff}\right)^{1/n}\right].
\end{equation}
Here $\rb$ is the core radius, $\Sigma_\mathrm{b}$ is the projected surface density at the core radius, $\gamma$ is the core slope, $\Reff$ is the effective (half-light) radius\footnote{As we assume a constant mass-to-light ratio, the half-light radius is equivalent to the projected half-mass radius.}, $n$ is the S\'ersic index, $b_n$ is estimated as $b_n\simeq 2n - 1/3 + 0.009876/n$ \citep{prugniel1997}, and $\alpha$ is the profile transition index.
We refer to the collective vector of these parameters as $\vb{\theta}\equiv[\rb, \Sigma_\mathrm{b}, \gamma, \Reff, n, \alpha]$.
We view the simulated merger remnant from $N_\mathrm{proj}=15$ random angles in $N_R=21$ logarithmically-spaced bins from $0.1\,\mathrm{kpc}$ to $30\,\mathrm{kpc}$, and use a Bayesian hierarchical model to fit the model parameters, as shown in \autoref{fig:dag}.
An excellent introduction to Bayesian data analysis, and in particular the strengths of hierarchical modelling in encoding dependencies between model parameters without overfitting, can be found in \citet{gelman2015}, which we summarise in \autoref{sec:app_fit} of the Appendix in the context of our work.
As part of the Bayesian inference workflow, we perform rigorous prior sensitivity analysis and posterior checking, also detailed in \autoref{sec:app_fit} of the Appendix.

\begin{table}
    \caption{Hyperparameter distributions for core-S\'ersic model. The distributions are either a normal ($\mathcal{N}$), exponential ($\mathrm{Exp}$), or gamma ($\mathrm{Gamma}$) distribution. The rationale for each distribution is discussed in \autoref{sec:app_fit} of the Appendix.}
    \label{tab:hyper}
    \begin{tabular}{llc}
        \hline
        Hyperparameter & Distribution & Truncation \\
        \hline
        $\mu_{\log_{10}\Sigma_\mathrm{b}}$ & $\mathcal{N}(10, 2)$ & None \\
        $\sigma_{\log_{10}\Sigma_\mathrm{b}}$ & $\mathcal{N}(0, 1)$ & $x>0$ \\
        $\lambda_\gamma$ & $\mathrm{Exp}(10)$ & None \\
        $\mu_n$ & $\mathcal{N}(8,4)$ & $0 < x \leq 15$ \\
        $\sigma_n$ & $\mathcal{N}(0,4)$ & $x>0$ \\
        $\sigma_\alpha$ & $\mathrm{Gamma}(2, 0.2)$ & None \\
        $\sigma_{r_\mathrm{b}}$ & $\mathcal{N}(0, 1)$ & $x>0$ \\
        $\sigma_{R_\mathrm{e}}$ & $\mathcal{N}(0, 12)$ & $x>0$ \\
        $\mu_{\tau_{\log_{10}\Sigma}}$ & $\mathcal{N}(0, 1)$ & $x>0$ \\
        $\sigma_{\tau_{\log_{10}\Sigma}}$ & $\mathcal{N}(0, 0.2)$ & $x>0$ \\
        \hline
    \end{tabular}
\end{table}

In \autoref{fig:density}, we see a clear decrease in the projected stellar mass density with increasing kick velocity for four representative values of $\vk$, consistent with the results in \autoref{fig:density3d}.
The shaded regions in \autoref{fig:density} correspond to the $25$ per cent Bayesian highest density interval (HDI), where the uncertainty is a direct result of choosing different projection angles when constructing the projected densities.

To discuss the difference in the evacuated core region of the merger remnants more quantitatively, we show the median (blue dashes) and interquartile range (IQR, blue boxes) of the core radius parameter $\rb$ from \autoref{eq:cs} for kick velocities $\vk\leq 1020\,\kmps$ in \autoref{fig:vkrb}, in units of both kpc and scaled to the $\vk=0\,\kmps$ core radius $\rbbin$, which corresponds to the pre-merger core radius of the galaxy.
Our results indicate that a non-zero kick velocity where the SMBH returns to the centre of the merger remnant serves to enlarge the core region relative to the core formed during the SMBH binary scouring phase.
In agreement with previous studies, we find a monotonic increase in the core radius $\rb$ with kick velocity.
Additionally, we see that the IQR of $\rb$ increases with increasing $\vk$, indicating that with an increased core size, the uncertainty introduced due to projection effects also increases.
To test the significance of the difference in distributions of $\rb$ between the $\vk=0\,\kmps$ case and the $\vk=1020\,\kmps$ case, we calculate the probability of superiority \citep[PS,][]{grissom2001} as:
\begin{equation}
    \mathrm{PS} = \frac{1}{n} \sum_i \mathbb{I}_{\rbbin^{(i)} < r_{\mathrm{b},1020}^{(i)}}(i),
\end{equation}
where the index $i$ runs over the $n$ posterior samples, and the indicator function $\mathbb{I}_A$ maps the $i^\mathrm{th}$ posterior draw to 1 if it belongs to the subset $A$, and to 0 otherwise.
With this measure we find that in more than 87 per cent of the posterior draws, $\rb$ from the $\vk=1020\,\kmps$ case is greater than $\rb$ from the $\vk=0\,\kmps$ case; the distributions are distinct.

As shown in the top right panel of \autoref{fig:csparams} in the Appendix, the projected density at the core radius decreases with increasing kick velocity, in agreement with the decreasing core density observed in the three dimensional density profiles in \autoref{fig:density3d}.
Similarly, the S\'ersic index $n$ also decreases with increasing kick velocity (bottom right panel of \autoref{fig:csparams}).
For recoil velocities $\vk\leq 300\,\kmps$, $n\simeq2.1$, and for $\vk\geq 660\,\kmps$, $n\simeq 1.6$.
For recoil velocities $360\,\kmps \leq \vk \leq 600\,\kmps$, the decrease in $n$ appears linear.
We note that the recoil velocity for which $n$ begins to decrease coincides closely with the stellar velocity dispersion of the SMBH binary-scoured core, $\sigma_{\star,0} \sim 270\,\kmps$.
The variation in $n$ can be understood in the context of the SMBH leaving the core, and thus affecting the stellar mass distribution beyond the core as it oscillates back to its equilibrium state.
Conversely, the effective radius $\Reff$ appears insensitive to the SMBH recoil velocity.

For the inner slope parameter $\gamma$, a weak trend to lower values at higher recoil velocities is apparent, after peaking at $\vk\sim 300\,\kmps \sim \sigma_{\star,0}$, though the distributions for $\gamma$ are broad.
Recently, \citet{khonji2024} demonstrated that increasing SMBH recoil velocity resulted in a suppressed $\gamma$, however the authors did not capture the $\vk \sim \sigma_{\star,0}$ regime of recoil velocities.
We understand the broad distributions that we observe for $\gamma$ as arising from the collapse of a three-dimensional density distribution to a two-dimensional representation, and then enforcing a spherically-symmetric one-dimensional profile to describe the two-dimensional surface density.
This is supported by the IQR of the $\gamma$ parameter remaining relatively broad with irrespective of kick velocity, unlike the core radius $\rb$ for example.

To investigate this further, for a subset of simulations we take particles in the centre of the merger remnant within a spherical annulus defined by $0.9\,\mathrm{kpc} \leq r \leq 1.0\,\mathrm{kpc}$ and consider sixty different, random projections into the two-dimensional plane. 
We perform a singular value decomposition (SVD) to obtain an ellipse representation of the projected annulus, and determine the ellipticity $\varepsilon$ of the projection
\begin{equation}
    \varepsilon = \frac{a-b}{a},
\end{equation}
where $a$ and $b$ are the semimajor and semiminor axes of the ellipse, respectively.
A value of $\varepsilon \neq 0$ indicates a departure from circular symmetry.
We find the distribution of $\varepsilon$ is positively-constrained and peaks at  $\varepsilon \simeq 0$ but is right-skewed, with the $\ordinal{75}$ percentile of $\varepsilon$ typically about $\sim0.05$.
This indicates that asymmetries are present in the projected stellar mass density profiles.
Non-circularity of the core region complicates using a radially-dependent density profile parameterised by a constant exponent $\gamma$: a range of possible values for $\gamma$ could describe the core region comparatively well.
The difficulty in constraining $\gamma$ provides good motivation for determining distributions for the parameters in \autoref{eq:cs} with Bayesian techniques, as opposed to point estimates achieved through typical frequentist methods.

\subsection{Estimating the amount of missing stellar mass}\label{ssec:missmass}

We continue the quantitative analysis of the core region by calculating the missing stellar mass of the merger remnants. 
As well as the core radius $\rb$, the mass deficit $M_{\mathrm{def}}$ can also be used to measure the core size. Past studies have found the mass deficit to correlate with the combined mass of the SMBH binary and the number of major `dry' mergers \citep[e.g.][]{milosavljevic2001, graham2003, merritt2006,rantala2024}.
In order to determine the mass displaced by the gravitational interactions and recoil, we adopt an observational approach by comparing the stellar mass density profiles of the inward extrapolated S\'ersic $\Sigma_{\mathrm{s}}(R)$ \citep{Sersic1963,Sersic1968} and the core-S\'ersic $\Sigma(R)$ fits  \citep[e.g.][]{dullo2014,bonfini2016, dullo2019, nasim2021b} as:
\begin{equation}
    M_{\mathrm{def}}=2\pi \int_0^\infty \left[ \Sigma_{\mathrm{s}}(R)-\Sigma(R)\right] R \dd{R},
\end{equation}
assuming a constant mass-to-light ratio \citep{bonfini2016}. 
The core-S\'ersic profile is calculated using \autoref{eq:cs} and \autoref{eq:cs_prefactor}, while the S\'ersic profile can be obtained using the same set of equations with a slight modification. 
By setting $\rb = 0$ in \autoref{eq:cs}, the profile becomes the standard S\'ersic profile:
\begin{equation}
    \Sigma_{\mathrm{s}} = \Sigma ' \exp \left( -b_n \left(\frac{R}{R_{\mathrm{e}}} \right)^{1/n} \right).
\end{equation}
To achieve the right normalisation, it is important to let $\rb$ retain its value in \autoref{eq:cs_prefactor} when constructing the S\'ersic profile \citep{graham2003}. 
In fitting the luminosity profiles, we perform for each kick velocity $10^4$ Monte Carlo samplings of the posterior distributions of the parameters $\vb{\theta}\equiv[\rb, \Sigma_\mathrm{b}, \gamma, \Reff, n, \alpha]$ described earlier in \autoref{ssec:projdens}, hence giving us $10^4$ realisations of the quantity $M_\mathrm{def}$.

The resulting mass deficits as a function of kick velocities $\vk\leq 1020\,\kmps$ are shown in the lower panel of \autoref{fig:vkrb}. At low kick velocities, $\vk\leq 240\,\kmps$, we see a slow monotonic increase of mass deficit as a function of the kick velocity. Although the scatter grows at higher kick velocities, an increase of the median trend can be observed over the whole kick velocity range.
For high $\vk$ we see a median trend of $M_\mathrm{def} \sim 4 M_{\mathrm{def},0}$, where $M_{\mathrm{def},0}$ corresponds to the mass deficit from binary scouring alone. 
We find that the median mass deficit due to SMBH binary scouring prior to the coalescence of the SMBHs is $M_{\mathrm{def},0}=1.46\times10^9\,\Msun$, roughly $0.25M_\bullet$.
As a consequence of the mass deficit growing to $M_\mathrm{def} \sim M_\bullet$ for non-zero recoil velocities, as well as $M_\mathrm{def}$ correlating positively with recoil velocity, we confirm our earlier hypothesis in \autoref{ssec:movement} that stellar mass bound to the SMBH (at most $0.37M_\bullet$) is not the primary mechanism by which the core grows.

There are numerous ways to calculate the mass deficit in simulations, and differences between methods along with varying initial conditions may contribute to discrepancies across studies. A common approach to calculating the mass deficit, which we also adopt, is to compare the S\'ersic and the extrapolated core-S\'ersic profiles \citep{graham2003, Dosopoulou2021, nasim2021b}. This technique has been used by \citet{dullo2014} for observations of cored galaxies, resulting in typical mass deficits of $0.5 \lesssim M_{\mathrm{def}} /M_\bullet \lesssim 4$; these mass deficits may not be due solely to SMBH binary scouring however. Other methods, including calculating the difference in densities between the initial and final states of the galaxy merger \citep[e.g.][]{merritt2006, gualandris2008}, and running the simulations with and without SMBHs \citep[e.g.][]{partmann2024, rantala2024}, typically find greater mass deficits than we do, but of the same magnitude.

As for core sizes presented in \autoref{fig:vkrb}, the interquartile ranges of the missing mass broaden with increasing kick velocities. This is due to growing uncertainties in the projection effects, even though the parameters $\vb{\theta}$ from the core-S\'ersic fits are Monte Carlo sampled as discussed at the end of \autoref{ssec:projdens}. Also, we have found a positive correlation between mass deficit and the core radius. Thus, the increasingly large range of $\rb$ values introduce a wider range of mass deficits as a function of kick velocity. 

\subsection{Predicted core size distribution}\label{ssec:predcore}
To predict the distribution of expected stellar core sizes given our merger remnant of two massive elliptical galaxies, we require a mapping from the kick velocity to the core radius. 
Let us define a scaled kick velocity $v'=\vk/v_\mathrm{esc}$.
Previous work by \citet{nasim2021b} fit an empirical power law of the form:
\begin{equation}\label{eq:vkrb_exp}
    \frac{\rb}{\rbbin} = K v'^q + 1,
\end{equation}
where $K$ and $q$ are free parameters.
\autoref{eq:vkrb_exp} gives that the core radius monotonically increases irrespective of kick velocity, na\"ively implying that as $\vk\rightarrow\infty$, $\rb\rightarrow\infty$.

We test two other empirical relations between $\rb$ and $\vk$.
The first is a simple linear relation:
\begin{equation}\label{eq:vkrb_lin}
    \frac{\rb}{\rbbin} = K v' + 1,
\end{equation}
and the second takes the form of a sigmoid function:
\begin{equation}\label{eq:vkrb_sig}
    \frac{\rb}{\rbbin} = K \left( 1 - e^{-q v'} \right) + 1.
\end{equation}

We fit each of the exponential, linear, and sigmoid models to the median $\rb$--$\vk$ trend in \autoref{fig:vkrb}, and display the relation with its best fit parameters in the same figure.
In the fit, we assume weakly informative positively-constrained normally distributed priors on the parameters, a normally distributed scatter $\tau$ truncated to $\tau>0$, and a Gaussian likelihood function.
The fit is found using the same Bayesian methods as in \autoref{ssec:projdens} (including posterior checking and prior sensitivity analysis), allowing for uncertainties to be estimated for the model fits.
The medians and 68 per cent HDI interval of the relation parameters in \autoref{eq:vkrb_exp}-\autoref{eq:vkrb_sig} are given in \autoref{tab:vkrb}.

\begin{table}
    \caption{Values of the fitted parameters in the three core radius -- kick velocity relations.}
    \label{tab:vkrb}
    \begin{tabular}{llcc}
        \hline
        Model & Parameter & Median & 68\% HDI \\
        \hline
        Exponential & $K$ & 2.72 & [2.19, 3.34] \\
        & $q$ & 0.78 & [0.57, 0.99] \\
        \hline
        Linear & $K$ & 3.03 & [2.39, 3.64] \\
        \hline
        Sigmoid & $K$ & 3.04 & [1.66, 5.73] \\
        & $q$ & 1.56 & [0.51, 3.07] \\
        \hline
    \end{tabular}
\end{table}

Using the three $\rb$--$\vk$ relations, we wish to construct the posterior predictive distributions for new data $\tilde{\rb}$ given the observed values of $\rb$:
\begin{equation}\label{eq:ppd}
    p(\tilde{\rb}|\rb) = \int p(\tilde{\rb} | \vb{\kappa}) p(\vb{\kappa} | \rb) \dd{\vb{\kappa}},
\end{equation}
where the integral marginalises out the respective $\rb$--$\vk$ relation parameters, collectively referred to as $\vb{\kappa}$ \citep{gelman2015}.
To generate the new data $\tilde{\rb}$, we first generate a distribution of kick velocities for our merger configuration by randomly sampling SMBH spin values, as we assume this is the only unconstrained variable in our simulated SMBH merger.
For the dimensionless SMBH spin parameter $\alpha_\bullet$ we use the distribution for dry mergers given in \citet{zlochower2015}, which takes the form of a beta distribution:
\begin{equation}
    P_\mathrm{Zlochower}(\alpha_\bullet) \propto (1-\alpha_\bullet)^{4.66884-1} \alpha_\bullet^{10.5868-1}
\end{equation}
and assume the direction of each SMBH spin vector is randomly distributed uniformly on the sphere.
With the distribution of $\alpha_\bullet$, and the state of the SMBH binary just prior to coalescence, we use the relations presented in \citet{zlochower2015} to obtain a distribution of SMBH recoil velocities.

\begin{figure}
    \centering
    \includegraphics[width=0.4\textwidth]{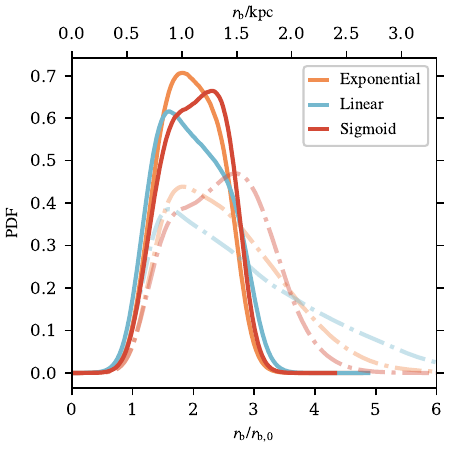}
    \caption{Probability density functions of transformation sampled core radii assuming a given model relation. 
    Solid lines depict a truncated recoil velocity distribution, whereas dash-dot lines depict a non-truncated recoil velocity distribution.
    The SMBH spin vectors are Monte Carlo sampled from the \citet{zlochower2015} relation assuming randomly-aligned azimuthal spins to generate a distribution of recoil velocities, which are then pushed through each of the three fitted models in \autoref{eq:vkrb_exp}, \autoref{eq:vkrb_lin}, and \autoref{eq:vkrb_sig}. 
    The sigmoid model is the preferred model, and has its mode at a larger value of $\rb$ than either the exponential or linear models, at $\rb/\rbbin\sim2.29$ ($\rb/\mathrm{kpc}\sim1.27$) in the truncated case.
    The (non-truncated) predicted kick velocities range from $\vk=0\,\kmps$ to $\vk\sim4000\,\kmps$.
    }
    \label{fig:rb_pdf}
\end{figure}

The new data $\tilde{\rb}$ is then generated by transformation sampling the distribution of kick velocities, and `pushing' the values through the desired $\rb$--$\vk$ relation.
Using transformation sampling allows us to obtain a distribution of core radii predicted by a given kick velocity model, as shown in \autoref{fig:rb_pdf}, whilst simultaneously accounting for uncertainty in the fit described by the $\rb$--$\vk$ relations.
Using $10^4$ Monte Carlo sampled values for $\vk$, where each value is used across 4 chains each of length 2000 samples, provides a total sample size of $8\times10^7$ values of $\tilde{\rb}$.

By only varying the SMBH spin vectors prior to merger, the range of kick velocities predicted by the model for the particular merger configuration in our study varies from $\vk=0\,\kmps$ to $\vkmax \equiv \vk\sim4000\,\kmps$, with a peak at $\vk \simeq 250\,\kmps$.
The maximum kick velocity that resulted in a settled SMBH in our sample, $\vk=1020\,\kmps$, corresponds to the $\sim\ordinalthird{63}$ percentile of the empirical cumulative distribution function (CDF) of the kick velocity distribution.
Additionally, from the empirical CDF we find that approximately 10 per cent of recoil velocities exceed the escape velocity of the centre of the merger remnant, $v_\mathrm{esc}=1800\,\kmps$.

The distribution of recoil velocities becomes slightly more right-skewed if the masses of the SMBHs are allowed to vary prior to merger.
By randomly sampling each SMBH mass as $M_{\bullet,\mathrm{new}} = 10^{\mathcal{N}(0, 0.3)} M_{\bullet,0}$, where the dispersion of 0.3 is the intrinsic scatter in the $M_\bullet$--$M_\mathrm{bulge}$ relation from \citet{haring2004}, we find that a recoil velocity of $1020\,\kmps$ corresponds to the $\sim\ordinal{76}$ percentile, and the central escape velocity of $v_\mathrm{esc}=1800\,\kmps$ corresponds to the $\sim\ordinal{95}$ percentile.
Thus, whilst a non-equal mass SMBH binary merger results in a slightly different recoil velocity distribution, large recoil velocities are still allowed, and are frequent, within the expected SMBH mass variation.
Herein we consider our fiducial case of equal-mass SMBH binaries, but do not expect the conclusions to drastically differ in the more general case.

As we cannot be guaranteed that the proposed $\rb$--$\vk$ relations hold beyond our fitted data range of $0\,\kmps$--$1020\,\kmps$, we discuss the distribution of core radii in two parts: the first limiting our analysis to $\vk \leq 1020\kmps$, and the second to $\vk \leq \vkmax$.

\subsubsection{Core size distribution for $\vk \leq 1020\kmps$}\label{sssec:coreT}
To truncate the kick velocity distribution with an upper bound of $1020\,\kmps$, we take only those values of $\vk$ which satisfy the velocity criterion, noting that as $1020\,\kmps$ corresponds to the $\sim\ordinalthird{63}$ percentile, we still have in excess of $4.5\times10^7$ samples of the core radius distribution.

We show the core radius distribution for the three different models with solid lines in \autoref{fig:rb_pdf}.
We find that the mode of the forward-folded core radius distribution is similar for the exponential and linear models, with $1.01\,\mathrm{kpc}$ ($1.82\,\rbbin$) and $0.88\,\mathrm{kpc}$ ($1.59\,\rbbin$), respectively.
Conversely, the mode for the sigmoid relation occurs at a larger value of $\rb$, namely $1.28\,\mathrm{kpc}$ ($2.31\,\rbbin$).

Looking at the shapes of the three distributions in \autoref{fig:rb_pdf}, we see a similarity in shapes between the exponential and linear models, which are both slightly right-skewed. 
This is due to the ever-increasing value of $\rb$ predicted for an increasing $\vk$: higher recoil velocities produce larger cores, but as higher recoil velocities become increasingly unlikely (assuming the \citet{zlochower2015} distribution), we observe the skew in the $\rb$ probability density function (PDF).
Conversely, the sigmoid model has a small hump prior to the mode, arising from the high likelihood of small kick velocities being pushed through the sigmoid function (\autoref{eq:vkrb_sig}) prior to the upper plateau of the function.
The peak in the $\rb$ distribution from the sigmoid model is a result of a large range of kick velocities (albeit with decreasing frequency) being mapped to the upper plateau of the sigmoid function, hence the correspondence between the median of the scale factor $K=3.04$ in \autoref{eq:vkrb_sig} and the distribution mode for the model, $2.31\,\rbbin$.

To determine which model of the three (exponential, linear, or sigmoid) best describes the data, we use approximate leave-one-out cross validation (LOO-CV) with Pareto smoothed importance sampling \citep[PSIS,][]{vehtari2017,vehtari2024}.
PSIS LOO-CV estimates the expected log pointwise predictive density ($\widehat{\mathrm{ELPD}}$), defined
\begin{equation}\label{eq:elpd}
    \widehat{\mathrm{ELPD}} = \sum_{i=1}^N \log \int p(r_{\mathrm{b},i} | \vb{\kappa}) p(\vb{\kappa} | r_{\mathrm{b},-i}) \dd{\vb{\kappa}},
\end{equation}
to determine the out-of-sample predictive accuracy of each model.
The integral in \autoref{eq:elpd} is analogous to \autoref{eq:ppd}, and gives the ability to predict the $i^\mathrm{th}$ data point $r_{\mathrm{b},i}$ when that same data point is excluded from the estimate of $\vb{\kappa}$, namely $p(\vb{\kappa} | r_{\mathrm{b},-i})$ \citep{gelman2014}.
A larger value of $\widehat{\mathrm{ELPD}}$ indicates better predictive ability of a given model compared to some other model \citep{vehtari2017,riha2024}.
When performing the LOO-CV, we ensure that the Pareto-$k$ values in the PSIS method are below 0.5, indicating reliable importance sampling estimates \citep{vehtari2024}.

We report the preferred model to be the sigmoid model (\autoref{eq:vkrb_sig}), with an $\widehat{\mathrm{ELPD}}$ of 1.70, compared to 1.58 and 0.61 for the exponential and linear models, respectively. However, we note that the difference in $\widehat{\mathrm{ELPD}}$ between the sigmoid and exponential models is not large, and the standard error of the difference in $\widehat{\mathrm{ELPD}}$, denoted $\widehat{\mathrm{SE}}\left( \Delta\widehat{\mathrm{ELPD}} \right)$, is 2.62 when comparing the sigmoid model to the exponential model, and 2.71 when comparing the sigmoid model to the linear model. This indicates that the difference between the models is likely not substantial in this range of recoil velocities.

\subsubsection{Core size distribution for $\vk \leq \vkmax$}\label{sssec:coreNT}
We show the core radius distribution, without a truncation in the recoil velocity distribution, for the three different models with dash-dot lines in \autoref{fig:rb_pdf}.
For the exponential and linear models, the distributions of $\rb$ are significantly right-skewed and broadened compared to when a truncation is applied to the recoil velocities.
The right-skew arises due to the derivatives of these models being positive and greater than unity, resulting in increasingly-higher recoil velocities being mapped to increasingly-larger core radii.
It is important to note that the mode of the $\rb$ distributions for the exponential and linear models with and without recoil velocity truncation are consistent, despite the differing shape of the distributions.
This is a result of low recoil velocities being more likely to occur, and hence contributing more to the transformation sampling of $\vk$ to $\rb$, than the high recoil velocity samples.

Similarly, the sigmoid model demonstrates a broader distribution of core radii when no truncation is applied to the recoil velocities compared to when a truncation is applied.
However the overall shape is more consistent with the truncated case than in the instance of the exponential or linear models.
The mode of the $\rb$ distribution is shifted to $1.49\,\mathrm{kpc}$ ($2.70\,\rbbin$) from $1.28\,\mathrm{kpc}$.
By construct, the sigmoid function is not as sensitive to large values of $\vk$ (compared to the exponential and linear models) if the function plateau is constrained to recoil velocities within our sampled range, as $\vk \rightarrow \rb$ and $\vk+\delta_{v_\mathrm{kick}} \rightarrow \sim \rb$.

Despite the three models mapping SMBH recoil velocities $0\leq \vk/\kmps \leq 1020$ to $\rb$ with a similar level of accuracy, as revealed by the $\widehat{\mathrm{ELPD}}$ measure, we argue that the sigmoid model is the overall preferred model when considering an unconstrained SMBH recoil velocity distribution.
The response of the stellar system to SMBH recoil velocities greater than the escape velocity of the system should be consistent, irrespective of how much greater $\vk$ is than $v_\mathrm{esc}$.
This behaviour is not captured with the exponential nor linear models.
For example, in the case of the sigmoid model, a scaled recoil velocity ($v'=v/v_\mathrm{esc}$) of $v'=2 \rightarrow \rb=3.9$, whereas $v'=4 \rightarrow \rb=4.0$.
Conversely, for the exponential model, $v'=2 \rightarrow \rb=5.7$, whereas $v'=4 \rightarrow \rb=9.0$.
From this argument, we suggest the sigmoid model to be the preferred model when considering the core radius produced by arbitrary SMBH recoil velocity.

Irrespective of using the full distribution of SMBH recoil velocities predicted by the \citet{zlochower2015} model, or a truncated distribution thereof, a recoiling SMBH serves to enhance the stellar core formed by the SMBH binary prior to merger typically by a factor of 2-3.
This leads to core sizes in excess of a kiloparsec, which is also seen in observations \citep[e.g.][]{thomas2016,dullo2019}.

\section{Orbit analysis}\label{sec:orbits}

\begin{figure*}
    \centering
    \includegraphics[width=\textwidth]{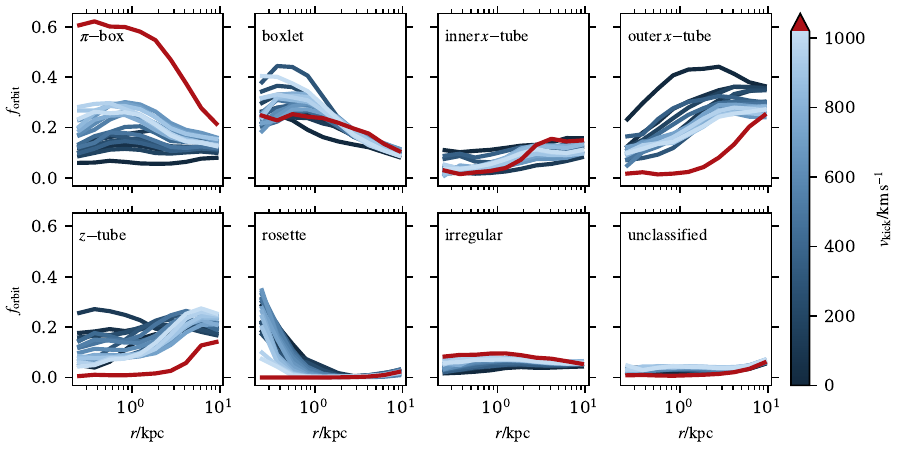}
    \caption{
        Orbit analysis of the merger remnants, with a consistent colour scheme to \autoref{fig:lagrangian}.
        Stellar particles are assigned to one of eight different orbital families, and binned into ten logarithmically-spaced radial shells such that $0.2 \leq R/\mathrm{kpc} \leq 11.0$.
        Noticeably, increasing recoil velocity induces a higher fraction of $\pi$-box orbits at all radii, and a complimentary decrease in rosette orbits at radii $r\lesssim1\,\mathrm{kpc}$.
    }
    \label{fig:orbits}
\end{figure*}

\begin{figure*}
    \centering
    \includegraphics[width=\textwidth]{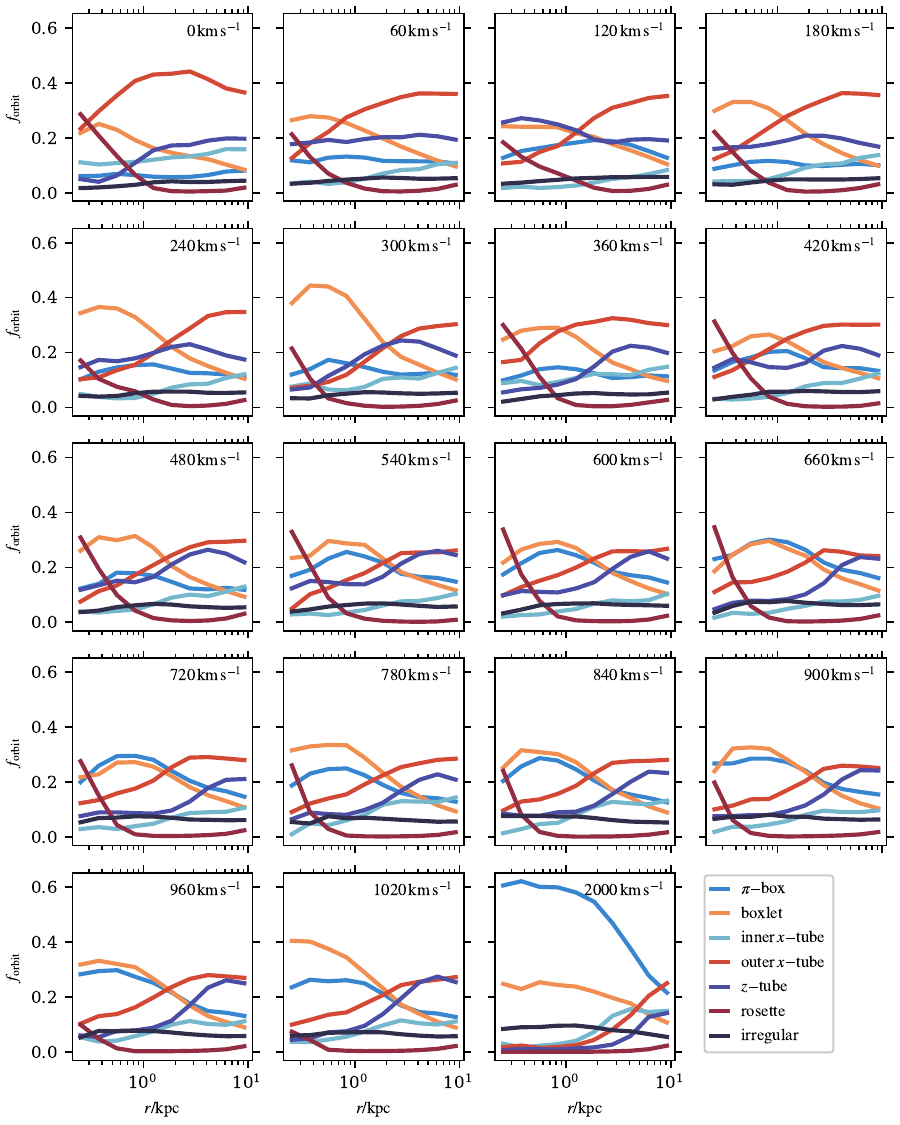}
    \caption{
        Same as \autoref{fig:orbits}, but each merger remnant is shown in its own panel, with the different orbital families depicted by colour, to better demonstrate the radial fraction of orbits for a given merger.
        Note that the orbits that were unable to be classified are not shown in this plot.
    }
    \label{fig:orbits2}
\end{figure*}

\begin{figure}
    \centering
    \includegraphics[width=0.48\textwidth]{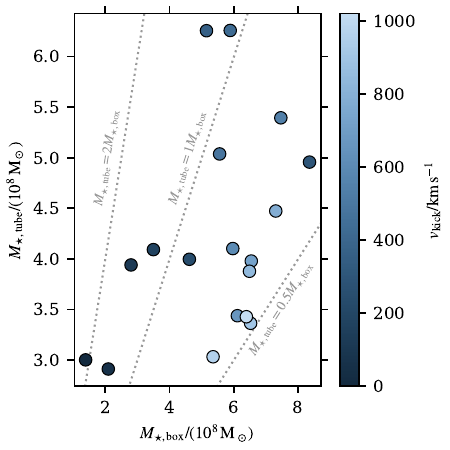}
    \caption{
        Amount of stellar mass classified as box orbits (boxlet or $\pi$-box) compared to tube orbits ($x$-tubes, $z$-tube, or rosette) within the median core radius $\rb$ for each simulation.
        Points are coloured by kick velocity.
        Simulations with low $\vk$ predominantly contain tube orbits within $\rb$, whereas larger recoil velocities have a tendency for box orbits to dominate. 
    }
    \label{fig:orbitsbt}
\end{figure}

Having established from a forward-modelling point of view that a changing SMBH recoil velocity induces a systematic effect in the mass distribution of the galaxy remnant, we turn our attention to inferring the occurrence of an SMBH that has experienced a recoil velocity, based on the snapshot when the SMBH is seen to be settled.
To this end, we perform an orbit analysis of the merger remnants in our sample following \citet{frigo2021}, which is briefly described here.

First, the merger remnant is rotated so that the $z$-axis coincides with the minor axis of the reduced inertia tensor (\autoref{eq:inertia}) as measured using the top 50 per cent most bound stellar particles to the stellar centre of mass.
We then need to fit the gravitational potential of the system, which we do so by using a self-consistent field (SCF) potential \citep{hernquist1992,jesseit2005,rottgers2014}, 
representing the potential $\Phi(\vb{x})$ generated by the particles in the simulation as a combination of harmonics with expansion coefficients $A$:
\begin{equation}
    \Phi(\vb{x}) = \sum_{nlm} A_{nlm} \Phi_{nlm}(\vb{x}),
\end{equation}
where each individual harmonic independently satisfies the Poisson equation.
The coefficients $A_{nlm}$ describe the order $n$ the radial component of the potential is expanded to, and to which order $l$, $m$ the angular components are expanded; the angular dependence is a (somewhat complicated) function of the spherical harmonics $Y_{lm}(\theta, \phi)$, where $|m|<l$ as usual.
The zeroth order of the potential expansion $\Phi_{000}(\vb{x})$ is the model upon which we wish to construct our expansion, and a common choice for this is the \citet{hernquist1990} profile.
Following \citet{frigo2021}, the expansion is limited to $n_\mathrm{max}=18$ and $l_\mathrm{max}=7$, as this captures the potential of the galactic merger remnant with sufficient accuracy.

The chosen basis functions are unable to describe a point-mass potential, thus complicating the simulations for which the SMBH has settled at the centre. 
To overcome this, we centre the snapshot on the SMBH position, but exclude the SMBH from the SCF routine.
This provides us with a potential field without the SMBH contribution; the potential from the SMBH is then added as a point-mass potential following the SCF potential construction.
In the case where the SMBH has not settled at the centre (i.e. the $\vk=2000\,\kmps$ simulation) we simply align the snapshot with the centre of mass determined using the shrinking sphere method on stellar particles, and neglect the contribution from the SMBH (as it is very far from the centre and thus no longer influences the orbits of the stellar particles).
We now have a representation of the potential in an analytical form in which the orbits of individual stellar particles can be integrated.
The potential of the merger remnant is checked for stability by comparing the potential from the SCF method to the potential computed from the particle data by \gadget{} during the simulation, and ensuring the ratio of the two is $\sim1$ at all radii.

Each stellar particle within $2\Reff \simeq 11\,\mathrm{kpc}$ of the centre is integrated for fifty orbits to determine which, if any, orbital resonances exist\footnote{As each stellar particle is integrated in a fixed potential, the effect of two-body interactions are not captured by this method.}.
The orbital resonances define the different families of orbits, as given in \citet{frigo2021}, and are listed in \autoref{tab:orbits}.
We then determine the fraction of each orbital family in ten logarithmically-spaced radial bins extending from $0.2\,\mathrm{kpc}$ to $11\,\mathrm{kpc}$.

\begin{table}
    \caption{Classification of orbital families}
    \label{tab:orbits}
    \begin{tabular*}{0.48\textwidth}{p{0.1\textwidth}p{0.35\textwidth}}
        \hline
        Family & Description \\
        \hline
        inner $x$-tube & Rotate about the major axis of the galaxy, but move radially when far from centre (concave shaped) \\
        outer $x$-tube & Rotate about the major axis of the galaxy (convex shaped) \\
        $z$-tube & Rotate about the minor axis of the galaxy \\
        $\pi$-box & Non-resonant motion with no net angular momentum (radial motion) \\
        boxlet & Resonant motion with no net angular momentum \\
        rosette & Typical orbit in a point-mass dominated spherically-symmetric potential \\
        irregular & No integrals of motion \\
        unclassified & Orbits unable to be classified \\
        \hline
    \end{tabular*}
\end{table}

Three distinct trends are made apparent in \autoref{fig:orbits}.
The first is the increase in the fraction of $\pi$-box orbits with increasing recoil velocity, most notable at radii $r\lesssim5\,\mathrm{kpc}$.
The simulation with $\vk=2000\,\kmps$ noticeably stands out, with the fraction of $\pi$-box orbits being $\sim0.6$ at the innermost radii. 
Complimentary but converse to the first trend, increasing the SMBH recoil velocity marginally decreases the fraction of $z$-tube orbits from $\sim 0.2$ to $\lesssim 0.1$ within $1\,\mathrm{kpc}$.
Again the $\vk=2000\,\kmps$ is noticeably offset from the other simulations, however remains consistent with the general trend.
Finally, increasing the SMBH recoil velocity decreases the fraction of rosette orbits at small radii $r<1\,\mathrm{kpc}$, with no rosette orbits for the $\vk=2000\,\kmps$ simulation.
There is evidence for a minor decrease in outer $x$-tube orbits with increasing recoil velocity at radii $r<3\,\mathrm{kpc}$, however this trend is not as dramatic as the three previously discussed.
In particular, the $\vk=2000\,\kmps$ simulation has almost no outer $x$-tube orbits until $r=2\,\mathrm{kpc}$, and only comes to dominate the orbit fraction at $r\sim10\,\mathrm{kpc}$.

We can naturally explain the decrease in rosette orbits with increasing recoil velocity by considering that a rosette orbit requires a point-mass like potential to orbit in.
With a larger recoil velocity, we are displacing the SMBH increasingly further from the stellar centre of mass, thus disrupting the conditions required to maintain such an orbit.
A consequence of disrupting these regular, rosette orbits, is the inducement of more radial orbits as recoil velocity grows, manifested as an increase in non-resonant $\pi$-box orbits. 
With a settled SMBH dominating the potential at the centre of the merger remnant, stellar orbits may slowly diffuse through relaxation processes to form rosette orbits at times $t>t_\mathrm{settle}$.
However, the timescale required for boxlet and $\pi$-box orbits to acquire sufficient angular momentum to become rosette orbits is very long; as discussed in \autoref{ssec:relax}, relaxation processes do not dominate the centre of the galaxy over the timescales relevant for these simulations, and violent relaxation is not applicable in the instance of a single SMBH.

We show the orbit families as a function of radius organised by recoil velocity in \autoref{fig:orbits2}, instead of family class.
At intermediate radii ($r>2\,\mathrm{kpc}$), outer $x$-tube orbits dominate over all other families for all simulations, a distinct feature of prolate systems.
The merger remnants in this study are predominantly prolate rotators at the time of analysis (see bottom panel of \autoref{fig:triax}), in agreement with results from previous simulation studies of major mergers \citep[e.g.][]{naab2003, gonzalez2009,ebrova2017}.
Interestingly, increasing the recoil velocity pushes the intersection of boxlet and $z$-tube orbits to increasingly larger radii, from $\sim 1.5\,\mathrm{kpc}$ for $\vk=0\,\kmps$ to $\sim 3\,\mathrm{kpc}$ for $\vk=1020\,\kmps$.
Finally, for the $\vk=2000\,\kmps$ simulation, the dominance of $\pi$-box orbits over all other families is apparent for radii $r\lesssim8\,\mathrm{kpc}$.
This is the only recoil velocity for which $\pi$-box orbits are the dominant orbital family; in the simulations with $\vk\leq1020\,\kmps$, $\pi$-box orbits are always subdominant to boxlet orbits.

Further insight is provided by grouping the orbital families into two broad classes, `boxes' and `tubes'.
Those orbits which do not have a net angular momentum we collectively refer to as boxes, and include boxlet and $\pi$-box orbits.
Similarly, those orbits which do have some net angular momentum we collectively refer to as tubes, and include the two types of $x$-tubes, $z$-tubes, and rosette orbits.
The amount of stellar mass within the median core radius $\rb$ for these two broad classes is shown in \autoref{fig:orbitsbt}, where the points are coloured by the SMBH recoil velocity.

Immediately we see that for low kick velocities, $\vk\leq 180\,\kmps$, tube orbits dominate the merger remnant core.
In particular, the $\vk=0\,\kmps$ case has the greatest ratio of tube to box orbits within the core, $M_{\star,\mathrm{tube}}/M_{\star,\mathrm{box}} = 2.0$. 
This is in agreement with an SMBH binary preferentially ejecting box orbits which lie within the loss cone during the three-body scattering phase.

Increasing the SMBH recoil velocity, there is a strong tendency for box orbits to dominate the core region, in some cases contributing more than $6.0\times10^8\,\Msun$ of stellar mass ($\sim0.1 M_\bullet$), compared to some $3.5\times10^8\,\Msun$ provided by tube orbits.
The recoil velocity where the transition from a tube orbit-dominated core to a box orbit-dominated core occurs is around $\vk\sim240\,\kmps$, approximately the same recoil velocity that allows the kicked SMBH to exit the SMBH binary-scoured core (see \autoref{fig:infall}, upper panel).
This suggests that as the kicked SMBH ploughs through the stellar core whilst it is settling but before its velocity falls below the velocity dispersion of the core, the rapid change in potential disrupts the conditions for angular momentum-conserving orbits, altering the stellar particles to radial orbits instead.
For the highest kick velocities, $\vk > 900\,\kmps$, the amount of stellar mass in box orbits is almost twice that in tube orbits, as shown in \autoref{fig:orbitsbt}.

In summary, a recoiling SMBH exiting the core leaves an imprint in the stellar orbits as it settles.
Consequently, the stellar kinematics are also affected by the SMBH motion.

\section{Stellar kinematics}\label{sec:kinematics}
\subsection{Integral field unit kinematics}\label{ssec:ifu}

\begin{figure*}
    \centering
    \includegraphics[width=0.99\textwidth]{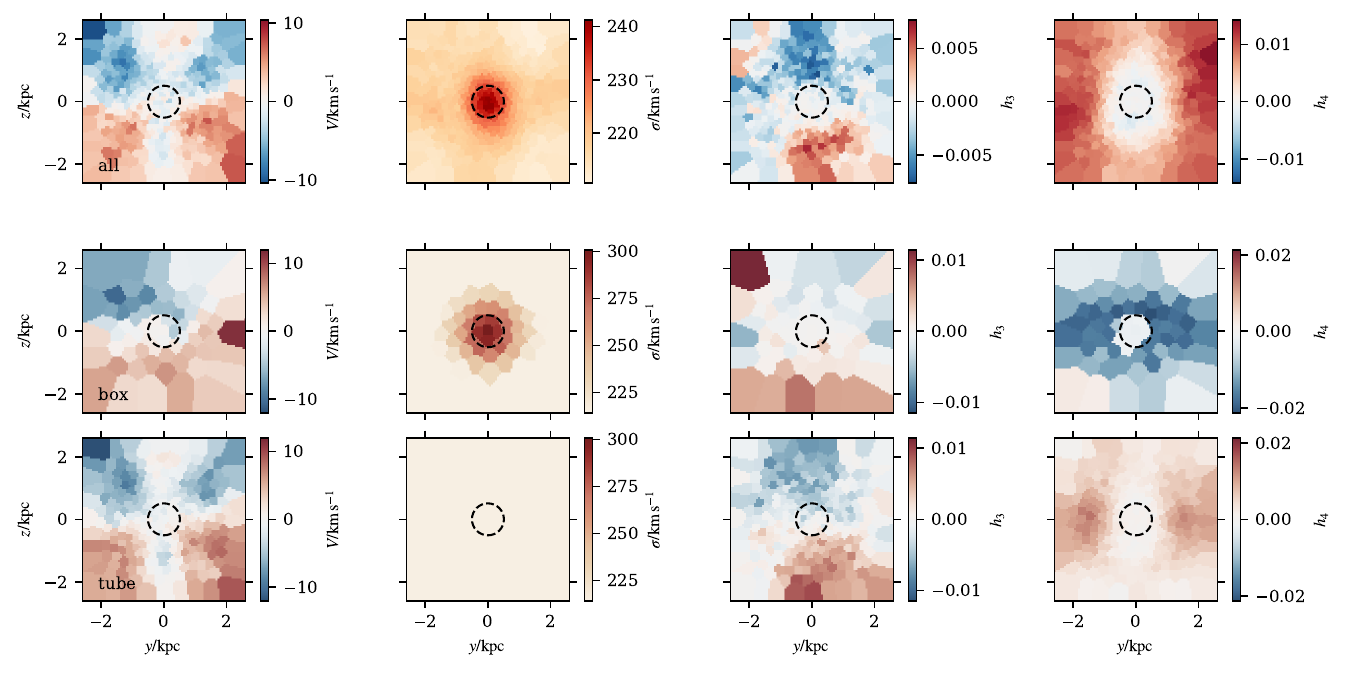}
    \caption{
        Mock integral field unit kinematic maps of the $\vk=0\,\kmps$ remnant stellar particles in the parallel projection at the time the SMBH has settled.
        The IFU maps extend out to $0.25\,r_{1/2}$ in a square aperture.
        The top row shows the IFU maps for each of the ordered rotation $V$, velocity dispersion $\sigma$, asymmetric deviation coefficient $h_3$, and symmetric deviation coefficient $h_4$, for all stellar particles within the square aperture.
        The second row shows the IFU maps for those stellar particles within the aperture classified as box orbits, and the third row shows the complementary IFU maps for those stellar particles within the aperture classified as tube orbits.
        The dashed circle indicates the core radius $\rb$ of the merger remnant.
        For the box and tube orbit rows, the colour scale is consistent for each column, however the colour scale is different for the top row.
        The ordered rotation $V$ is generally correlated with $h_3$.
        A centrally-peaked velocity dispersion is only visible in box orbits and not tube orbits, as is an annulus of negative $h_4$.
        Almost no negative $h_4$ is present in the IFU map for all stellar particles.
        From \autoref{fig:orbitsbt}, this simulation has $1.4\times10^8\,\Msun$ stellar mass in box orbits, and $3.0\times10^8\,\Msun$ stellar mass in tube orbits, within $\rb$.
    }
    \label{fig:IFU0000}
\end{figure*}

\begin{figure*}
    \centering
    \includegraphics[width=0.99\textwidth]{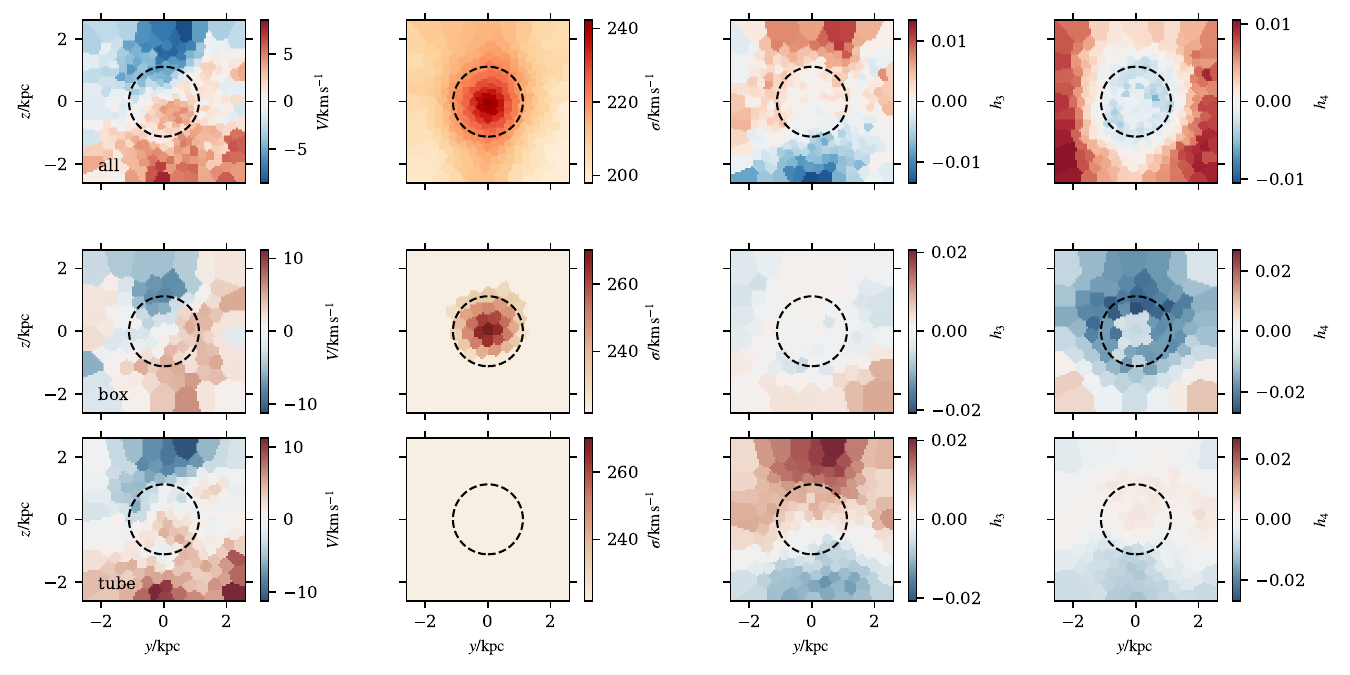}
    \caption{
        Similar to \autoref{fig:IFU0000}, mock integral field unit kinematic maps of the $\vk=600\,\kmps$ remnant stellar particles in the parallel projection at the time the SMBH has settled.
        From \autoref{fig:orbitsbt}, this simulation has $5.9\times10^8\,\Msun$ stellar mass in box orbits, and $4.1\times10^8\,\Msun$ stellar mass in tube orbits, within $\rb$.
        Similar to the $\vk=0\,\kmps$ case, a peaked velocity dispersion is visible for all stellar particles, but this contribution comes solely from the box orbits.
        An annulus of negative $h_4$ is seen to extend to the core radius in the IFU map for all stellar particles, and for box orbits.
    }
    \label{fig:IFU0600}
\end{figure*}

\begin{figure}
    \centering
    \includegraphics[width=0.5\textwidth]{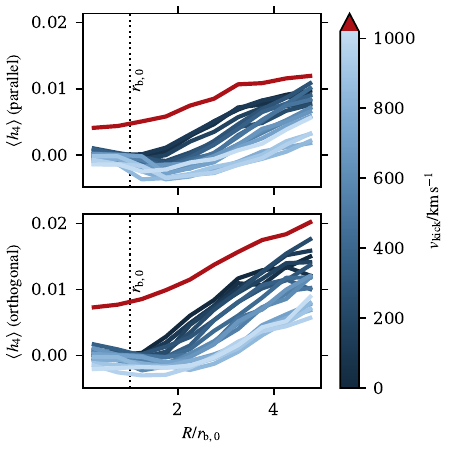}
    \caption{
        Annuli-averaged $h_4$ parameter for the parallel (top panel) and orthogonal (bottom panel) projections.
        For both the parallel and orthogonal projections, for those simulations with $\vk \leq 1020\,\kmps$ (i.e. have their SMBHs settled), there is a systematic decrease in the $h_4$ parameter with increasing $\vk$ for $\vk\lesssim 720 \,\kmps$.
        For $720 \lesssim\vk/\kmps\leq 1020$, the $h_4$ parameter is indistinguishable from each other. 
        For the $\vk=2000\,\kmps$ case, the $h_4$ is both positive and significantly larger than the $720\lesssim\vk/\kmps\leq1020$ simulations, for both parallel and orthogonal projections. 
    }
    \label{fig:h4}
\end{figure}

We next create mock integral field unit (IFU) observations using two different projections for the line-of-sight (LOS) velocity distribution: one where the LOS is along the SMBH kick axis (`parallel'), and another where the LOS is along an axis orthogonal to the SMBH kick axis (`orthogonal').
Testing these two special projections, we investigate the limiting cases of how the galaxy kinematics behave when all or none of the SMBH motion is directed along the LOS to the observer.

We create our mock IFU observations following \citet{naab2014}, assuming our merger remnants represent local elliptical galaxies that lie at $z=0.01$.
Our observations encompass a square aperture of side length $0.25r_{1/2}$, where $r_{1/2}$ is the 3D half mass radius. 
For each particle within this aperture, we generate 25 pseudo-particles with identical LOS velocities as the original particle, but are spatially displaced from the original particle with a Gaussian distribution in the projected $x$ and $y$ direction with a standard deviation of $6.8''$ (corresponding to a physical scale\footnote{\url{https://cosmocalc.icrar.org} with $\Lambda\mathrm{CDM}$ cosmological parameters $H_0=70\,\kmps\,\mathrm{Mpc}^{-1}$, $\Omega_\mathrm{M}=0.3$, and $\Omega_\Lambda=0.7$.} of $0.3\,\mathrm{kpc}$ at $z=0.01$) to mimic seeing effects.
The pseudo-particles are then binned onto a spatial grid, where the grid is assumed to have a resolution of $0.2''$, similar to the observations of \citet{neureiter2023}, which then corresponds to a physical resolution of $\sim0.04\,\mathrm{kpc}$.
Following the method in \citet{cappellari2003}, we then group adjacent bins into larger Voronoi bins to achieve a particle number of $\sim 50000$ particles per Voronoi bin.

For each Voronoi bin, we follow \citet{vandermarel1993} and decompose the LOS velocity into a series of Gauss-Hermite functions, described by the mean radial velocity $V$, the velocity dispersion $\sigma$, asymmetric deviation $h_3$, and symmetric deviation $h_4$.
The line profile is thus described by:
\begin{equation}
    \mathscr{L} = \frac{1}{\sqrt{2\pi}\sigma} e^{-w^2/2} \left\{ 1 + \sum_{j=3}^4 h_j H_j(w) \right\},
\end{equation}
where $w \equiv (v_\mathrm{LOS} - V)/\sigma$, and the Hermite polynomials $H_3$ and $H_4$ are defined:
\begin{align}
    H_3 &= \frac{1}{\sqrt{3}} (2w^3 - 3w) \nonumber \\
    H_4 &= \frac{1}{\sqrt{24}} (4w^4 - 12w^2 + 3).
\end{align}
Relative to a standard Gaussian (which has $h_4=0$), a positive value of $h_4$ corresponds to extended high velocity tails in the LOS velocity distribution and a more peaked distribution about the mean value $V$.
Conversely, a negative value of $h_4$ corresponds to weaker tails in the LOS velocity distribution and a broader distribution of LOS velocities about the mean value $V$.
A recent comprehensive study of LOS velocity distribution fitting applied to elliptical galaxies can be found in \citet{mehrgan2023}.

We show example IFU maps for the parallel projections of the $\vk=0\,\kmps$ case in \autoref{fig:IFU0000}, and of the $\vk=600\,\kmps$ case in \autoref{fig:IFU0600}.
In both figures, the first row corresponds to mock observations of all stellar particles within the square aperture, the second row corresponds to the subset of particles classified as box orbits, and the third row corresponds to the subset of particles classified as tube orbits.
The first column of each figure depicts the ordered rotation $V$ of the stellar particles, the second column shows the velocity dispersion $\sigma$, the third column shows the coefficient of asymmetric deviation $h_3$, and the fourth column shows the coefficient of symmetric deviation $h_4$.
The core radius of the merger remnant, as determined in \autoref{ssec:projdens}, is depicted with a dashed ring in each panel.

We first note that all our simulated merger remnants have almost negligible rotation, as immediately seen in the ordered rotation maps of \autoref{fig:IFU0000} and \autoref{fig:IFU0600}.
Quantitatively, the dimensionless spin parameter is a useful measure to assess the level of rotation \citep{emsellem2007}, and is defined:
\begin{equation}
    \lambda_R = \frac{\sum_{i=1}^N F_i R_i |V_i|}{\sum_{i=1}^N F_i R_i \sqrt{V_i^2 + \sigma_i^2}},
\end{equation}
where $F_i$ is the flux, $R_i$ is the radial displacement from the centre, $V_i$ is the mean velocity, and $\sigma_i$ is the velocity dispersion, each taken within the $\ordinal{i}$ Voronoi bin.
We find that our galaxies have $\lambda_R<0.02$ within the IFU aperture, and thus we conclude that very little rotation is present.

For the $\vk=0\,\kmps$ case in \autoref{fig:IFU0000}, we first note an interesting structure in the ordered rotation $V$ of all stellar particles. 
There is a hint of rotation occurring about the $z=0$ axis, however a counter-rotating structure is present close to the $y=0$ axis (keeping in mind that as previously discussed, overall the magnitude of this rotation is very minimal).
This feature is also seen in the mock IFU images of the tube orbits.
Additionally, creating mock IFU maps for two other projections (one with the line-of-sight aligned with the major axis of the inertia tensor, and the other with the line-of-sight aligned with the vector normal to the merger plane), this feature is not present. 
We thus conclude that immediately following the merger of an SMBH binary, complex structure in the LOSVD may be apparent for tube orbits.

The kinematic features for the second, third, and fourth moments of the LOSVD are not as sensitive to the viewing angle, with box orbits displaying a centrally-peaked velocity dispersion compared to a spatially homogeneous velocity dispersion for tube orbits.
The spatial extent of the centrally peaked velocity dispersion extends beyond the core radius when viewing all stellar particles, however is overall reduced in magnitude compared to the box orbits owing to the inclusion of the tube orbits.
A peaked central velocity dispersion can be naturally explained for box orbits owing to the radial motion of these particles increasing in velocity closer to the centre of the potential, when these particles are at their pericentre.

Another curious feature is an annulus of negative $h_4$ that is clearly visible for box orbits, but is absent in the $h_4$ maps for tube orbits and all stellar particles.
At the centre of the annulus, $h_4\sim0$, with the inner ring of the annulus coinciding with the core radius.
The region of negative $h_4$ in the IFU maps of the box orbits can be understood as the population of box orbits which did not interact with the SMBH binary prior to coalescence. 
The positive $h_4$ beyond the annulus is a result of the box orbits reaching apocentre, where the velocity becomes zero (irrespective of orientation), thus resulting in an overly peaked LOSVD.
From \autoref{fig:orbitsbt}, for $\vk=0\,\kmps$ there is more stellar mass in tube orbits than box orbits within the core (and at radii $r>\rb$, shown by the increased number of Voronoi cells for tube orbits in \autoref{fig:IFU0000} compared to box orbits).
More tube orbits leads to the $h_4$ annulus being absent in the map of $h_4$ for all stellar particles.

Turning our attention to the IFU maps in \autoref{fig:IFU0600} for the $\vk=600\,\kmps$ case, we notice immediately that ordered rotation about the $z=0$ axis is minimal, and the velocity dispersion for all stellar particles has a centrally-peaked structure that extends to the core radius.
Converse to the $\vk=0\,\kmps$ case the core radius extends to the same radius as the enhanced velocity dispersion in the box orbits.

Similar to the $\vk=0\,\kmps$ case, an annulus of negative $h_4$ is present for the $\vk=600\,\kmps$ simulation.
However, the core radius has expanded beyond the central $h_4\sim0$ region to encompass some of the annulus.
Owing to the box orbit-dominated core (from \autoref{fig:orbitsbt}, $M_{\star,\mathrm{tube}} / M_{\star,\mathrm{box}} \sim 0.75$), the annulus of negative $h_4$ is now present in the $h_4$ kinematic map for all stellar particles.
Critically however, a central region of $h_4\sim0$ is present in the map for all stellar particles, and corresponds to the same spatial extent of the $h_4\sim0$ region in the $\vk=0\,\kmps$ simulation. 
We propose that this central region of $h_4 \sim 0$ is indicative of the core radius due solely to SMBH binary scouring $\rbbin$, prior to any core enhancement by the recoiling SMBH, and set about identifying the radial extent of this region.

\subsection{Radial trends in $h_4$}
To quantify the radial dependence of $h_4$, we bin the Voronoi cells into ten concentric annuli out to $5 \rbbin$, and take the median $h_4$ value within each annulus, denoting this as $\langle h_4 \rangle$.

The radial dependence of $\langle h_4 \rangle$ is shown in \autoref{fig:h4}.
Aside from the $\vk=2000\,\kmps$ simulation, increasing the kick velocity results in progressively more negative $\langle h_4 \rangle$ in both the parallel and orthogonal projections at radii $R\gtrsim1\,\mathrm{kpc}$, seen as an overall downward shift in \autoref{fig:h4}. 
The value of $\langle h_4 \rangle$ saturates for simulations with $\vk\gtrsim 720\,\kmps$, with a minimum of $\langle h_4 \rangle < 0$ at $R \sim 1\,\mathrm{kpc}$ for both the parallel and orthogonal projections.

As discussed in \autoref{ssec:ifu}, identifying the radius at which $\langle h_4 \rangle$ starts to become negative can give insight to the core size due solely to SMBH binary scouring. 
For the parallel projection, $\langle h_4 \rangle$ starts to become negative at $R \sim 0.5\,\mathrm{kpc}$, and $R \sim 0.6\,\mathrm{kpc}$ for the orthogonal projection, irrespective of the recoil velocity $\vk$.
Both of these values are in good agreement with the known core size due to binary scouring of $0.58\,\mathrm{kpc}$ discussed in \autoref{ssec:settle}, as indicated by the vertical dotted line in \autoref{fig:h4}.

To understand the overall trends in \autoref{fig:h4}, we must refer to the orbit analysis of \autoref{sec:orbits}.
Within $R \sim 1\,\mathrm{kpc}$, the $\langle h_4 \rangle$ profiles are largely consistent between kick velocities.
The dominant orbital family within this radius for simulations with recoil velocities $60 \leq \vk/\kmps \leq 1020$ is the boxlet family: radial orbits which can reach arbitrarily close to the centre.
As these orbits approach their pericentre at small radii, the different projections of these boxlet orbits result in a LOSVD that is relatively broad around the mean, resulting in a value of $\langle h_4 \rangle$ that is either 0 or slightly negative, irrespective of the SMBH recoil velocity.
This also explains the consistency in $\langle h_4 \rangle$ between the parallel and orthogonal projections at small radii.

At radii beyond $R\sim1\,\mathrm{kpc}$, there is a strong tendency, at a given radius, for simulations with a low $\vk$ to have a higher value of $\langle h_4 \rangle$ than simulations with a high $\vk$.
From \autoref{fig:orbits2}, this is also the radius at which the dominant orbital family transitions from boxlet orbits to $x$-tube orbits for simulations with $60 \leq \vk/\kmps \leq 1020$, though boxlet orbits are by no means negligible.
Additionally, when considering the apocentre distribution of stellar particles with a mean position within $\sim0.25r_{1/2}$ (\autoref{fig:apo_dist}), the distributions are all left-skewed, and become increasingly left-skewed with increasing $\vk$.
At apocentre, a stellar particle on a radial orbit will have a zero velocity vector, which is then viewed as a zero velocity vector when projected to an observer.
Similarly, for stellar particles on non-radial orbits, the apocentre represents a minimum, but not necessarily zero, velocity\footnote{Except in the case where the motion is completely tangential to the observer, in which case there is no projected component of the velocity vector.}.
In either case, the LOSVD peaks towards the mean velocity at apocentre, thus inducing a mildly-positive $\langle h_4 \rangle$.
As the peak of the apocentre distribution shifts to larger radii with increasing $\vk$, we expect to find more stellar particles at apocentre at small radii for low $\vk$ simulations, than for high $\vk$ simulations.
This in turn leads to an elevated value of $\langle h_4 \rangle$ at small radii for simulations with low $\vk$ compared to those with high $\vk$.

For the $\vk=2000\,\kmps$ simulation, where the SMBH did not return to the centre of the merger remnant, the radial $\langle h_4 \rangle$ profile is clearly different to those simulations in which the SMBH managed to return to the centre.
Here, the $\langle h_4 \rangle$ profile is elevated compared to the lower $\vk$ simulations at all radii in both the parallel and orthogonal projections.
Again, we can readily understand this phenomenon with the orbital analysis in \autoref{sec:orbits}.
In \autoref{fig:orbits2}, we clearly see that $\pi$-box and boxlet orbits dominate over all other orbital families in the radial range covered by the mock IFU maps.
Additionally, the apocentre distribution of the $\vk=2000\,\kmps$ simulation peaks at smaller radii than the majority of the simulations with a settled SMBH: it is less left-skewed, indicating that more particles have their apocentre at radii covered by the mock IFU maps.
By the same reasoning as above, a higher concentration of apocentres at small radii narrows the LOSVD profile, resulting in an elevated value of $\langle h_4 \rangle$, and is seen in \autoref{fig:h4}.

In the analysis of the kinematic maps, a complication arises from choosing the snapshot once the SMBH has settled, which as discussed in \autoref{ssec:settle} varies as a function of recoil velocity.
To explore the effect of our choice of snapshot, we redo the IFU analysis for various times following when the SMBH kick occurs.
A representative example is shown in \autoref{fig:h4900} for the case of $\vk=900\,\kmps$.
We take snapshots corresponding to times between $0.18\,\mathrm{Gyr}$ and $1.70\,\mathrm{Gyr}$, in addition to the fiducial snapshot at $1.28\,\mathrm{Gyr}$.
Prior to the SMBH settling, the value of $\langle h_4 \rangle$ fluctuates as the settling SMBH influences the surrounding stellar kinematics.
The $\langle h_4 \rangle$ profile is consistent for all times following the SMBH settling, and thus gives us confidence that, provided the SMBH has settled, the $\langle h_4 \rangle$ profile is insensitive to the precise time of measurement.
As a result, the differences in $\langle h_4 \rangle$ profile presented in \autoref{fig:h4}
 are a result of the stellar mass distribution responding to the change in potential caused by the oscillating SMBH, and not due to general relaxation processes (refer also to \autoref{ssec:relax}).

\begin{figure}
    \centering
    \includegraphics[width=0.48\textwidth]{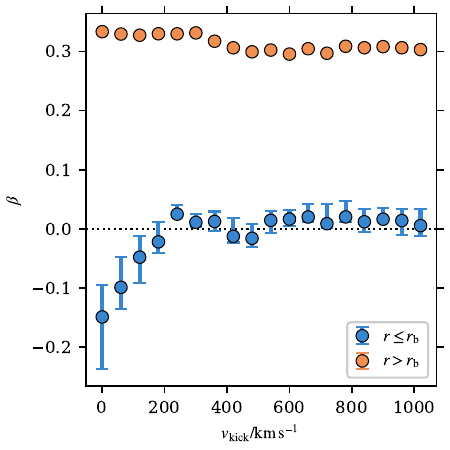}
    \caption{
        Stellar velocity anisotropy parameter $\beta$ as a function of the kick velocity $\vk$. 
        The stellar particles are divided into two radial bins with $r \leq \rb$ and $r>\rb$, where the core radius $\rb$ is sampled $N=10^4$ times for each simulation from its marginal $\rb$ posterior distribution.
        The final value for $\beta$ and the corresponding error are computed as the median and interquartile range across the entire sample (note for the orange points, the error is smaller than the marker). 
        The blue and orange points correspond to the inner and outer bins, respectively, while the dotted horizontal line indicates an ergodic system with $\beta=0$. 
        Only simulations with $\vk \leq 180 \, \kmps$ display tangential bias ($\beta < 0$) for $r \leq \rb$, whereas all simulations are radially-biased for $r> \rb$.
    }
    \label{fig:beta}
\end{figure}

\subsection{Velocity anisotropy}
As described in \autoref{sec:orbits}, SMBHs can induce a significant change in the radial distribution of the stellar orbital families. On a more global scale, the change in orbit distribution manifests as a stellar velocity anisotropy. A widely used measure for the velocity anisotropy of a given stellar system is the anisotropy parameter $\beta$ \citep{binney2008} given by
\begin{equation}
    \beta = 1 - \frac{\sigma^2_{\theta} + \sigma^2_{\phi}}{2 \sigma^2_r},
\end{equation}
where $\sigma_{\theta}$, $\sigma_{\phi}$, and $\sigma_{r}$ are the components of the stellar velocity dispersion in spherical coordinates. 
Stellar systems with $\beta < 0$ are referred to as `tangentially-biased' systems, whereas $\beta > 0$  corresponds to `radially-biased' systems. 
A stellar system with purely radial orbits ($\sigma_{\theta}=\sigma_{\phi}=0$) corresponds to $\beta = 1$, while purely circular orbits ($\sigma_{r}=0$) would yield $\beta = -\infty$.
An ergodic system is one where $\sigma_{\theta}=\sigma_{\phi}=\sigma_r \iff \beta=0$, and hence corresponds to a balance between box and tube orbits.

Prior to their eventual merger, SMBH binaries can interact strongly with the stellar particles in their vicinity via three-body scattering. 
However, in order to undergo a strong interaction, the stellar particle must come sufficiently close to the SMBH binary. 
Such stellar particles are predominantly on radial orbits with small pericentre distances and low angular momenta, which enables them to reach the centre of the galaxy: they belong to the loss cone of the SMBH binary. 
As a consequence, SMBH binaries are expected to induce tangentially-biased cores in the central regions of their host galaxies by primarily ejecting stellar particles on radial orbits \citep[e.g.][]{quinlan1997,milosavljevic2001,Thomas2014,rantala2018,rantala2019}. 
Recoiling SMBHs, on the other hand, can induce velocity anisotropy by changing the shape of the local gravitational potential as described in \autoref{sec:orbits}. Rosette orbits, for instance, are easily disturbed by a recoiling SMBH: as the potential in the initial position of the SMBH becomes non-spherical, stars initially on rosette orbits can move onto $\pi$-box orbits, which manifests as additional tangential bias in the central regions of the galaxy. 

\autoref{fig:beta} shows the anisotropy parameter $\beta$ as a function of $\vk$ for the merger simulations once the SMBH has settled, for $\vk$ between $0\,\kmps$ and $1020\,\kmps$.
For each simulation, the stellar particles are divided into two radial bins corresponding to $r \leq \rb$ (blue points) and $r>\rb$ (orange points), where the core radius $\rb$ is Monte Carlo sampled $N=10^4$ times for each simulation from its marginal $\rb$ posterior distribution (\autoref{ssec:projdens}).
The final $\beta$ and the corresponding error are obtained as the median and interquartile range across the entire sample for each merger remnant.

As is evident from \autoref{fig:beta}, the merger remnants with $\vk \leq 180 \, \kmps$ have tangentially-biased ($\beta<0$) stellar populations within $\rb$.
This is consistent with \autoref{fig:orbitsbt}, in which increasing the kick velocity is shown to induce a lower fraction of angular momentum-conserving orbits ($x$-tube, $z$-tube, and rosette) and a higher fraction of radial orbits ($\pi$-box and boxlet) in the inner regions.
For larger kick velocities ($\vk\gtrsim\sigma_{\star,0}$), the $\beta$ parameter for stellar particles within $\rb$ is consistent with an ergodic system. 
This results from the growth of the core radius with increasing $\vk$ shown in \autoref{fig:vkrb}: as $\rb$ increases, the enlarged core region encompasses stellar particles on box orbits which were not ejected by the SMBH binary prior to coalescence, as demonstrated by the IFU maps in \autoref{fig:IFU0600}. 
These stellar particles, found at radii $\rbbin < r < \rb$ (recalling that $\rbbin$ is the core radius due to SMBH binary scouring only), have apocentres that in general increase with increasing kick velocity. 
As the said stellar particles pass radially within the core region, they increase the range of radial velocities within $\rb$, leading to a larger radial velocity dispersion so that $2 \sigma_r^2 \simeq \sigma_\theta^2 + \sigma_\phi^2$ and $\beta \approx 0$ for the ergodic systems seen in \autoref{fig:beta}.
In this way, the anisotropy parameter $\beta$ of the core region of an elliptical galaxy might give insight into whether or not the recoiling SMBH was ejected with a $\vk$ that removed it from the SMBH binary-scoured core.

For $r>\rb$, all merger remnants display a strong radial bias of $\beta \gtrsim 0.3$. This might initially seem contradictory with \autoref{fig:orbits}, which shows that the tangentially-biased outer x-tube and z-tube orbits are the dominant orbital families at $r\sim10$ kpc. 
However, the stellar mass density decreases rapidly towards $r\sim10$ kpc as shown in \autoref{fig:density3d} and hence the outer regions contribute only weakly to the overall $\beta$ for $r>\rb$. 
Furthermore, the tangential bias of the aforementioned orbital families is weak at large radii, as shown in Figure 9 of \citet{frigo2021}, whereas the non-negligible $\pi$-box and boxlet orbits display strong radial bias at large radii. 
Finally, it is worth noting that the radial bias towards large radii is not induced by the recoiling SMBH: as is evident, the outer $\beta$ bins of all merger remnants with $\vk>0\,\kmps$ are consistent with the $\vk=0 \, \kmps$ remnant, which likewise displays a strong radial bias for $r > \rb$. 
Instead, the radial bias is primarily induced by the (radial) merger orbit, which enables stellar particles to be deposited on radially-biased orbits during the merger. A radially-biased outer region is also predicted by cosmological galaxy formation models, in which the late-stage growth of elliptical galaxies is attributed to numerous minor mergers \citep[e.g.][]{Naab2009}. As discussed in \citet{rantala2018}, further mergers would likely induce even stronger radial bias towards the outer regions and, assuming the inclusion of new SMBHs, also stronger tangential bias in the central regions.  

\section{Discussion} \label{sec:discussion}
\subsection{Comparison to literature}

\begin{figure}
    \centering
    \includegraphics[width=0.48\textwidth]{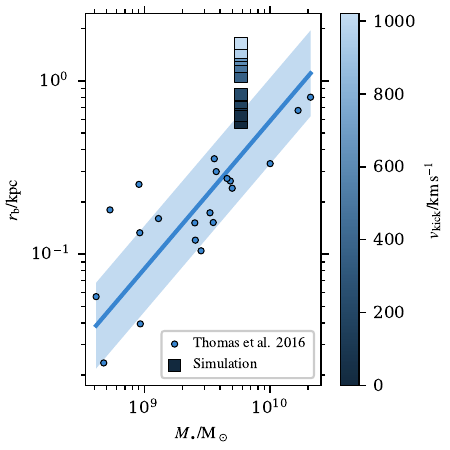}
    \caption{
        Comparison of the median measured core radii for each merger remnant to the observational data and the best fit constant-exponent power law presented in \citet{thomas2016}.
        The shaded region shows the reported intrinsic scatter of the relation.
        Only the merger remnants with very low $\vk$ lie within the intrinsic scatter of the relation, whereas remnants with large $\vk$ are significantly offset from the relation.
        }
    \label{fig:core-obs}
\end{figure}

In agreement with previous work \citep{boylan2004,merritt2004}, we find that an SMBH displaced due to gravitational recoil decreases the stellar density in the central regions of the merger remnant as it settles to an equilibrium state.
Additionally, we find that this oscillatory motion, by the time the SMBH has settled, increases the size of the stellar density core beyond that induced by the scouring of the SMBH binary prior to coalescence, confirming the results of \citet{gualandris2008}, \citet{nasim2021b}, and \citet{khonji2024}.
We assess the magnitude of this scouring with time using the Lagrangian radii presented in \autoref{fig:lagrangian}.
We see a sharp increase in the Lagrangian radii immediately after the SMBH binary coalesces.
If the SMBH is kicked beyond the core, the SMBH generally does not pass through the core during the next few orbits due to tangential motion induced by the stellar potential torquing the SMBH.
When the SMBH does pass through the core region in the last orbits when the SMBH velocity $v_\bullet$ is only marginally greater than the stellar velocity dispersion within the core, the Lagrangian radius rapidly increases further. 
\citet{gualandris2008} found that each oscillation of the SMBH enlarges the core, whereas \citet{nasim2021b} found that only the initial oscillation due to the recoil kick affects the core. 
Our result suggest that differences in the torquing due to the specific distribution of stellar mass could alleviate some of the tension between the two studies. 

The three-dimensional stellar density profile (\autoref{fig:density3d}) appears centrally-flat only in the instance where the SMBH escaped the galaxy entirely ($\vk=2000\,\kmps$). Conversely, a mild density cusp $\rho \propto r^{-\nu}$ with $0 < \nu < 1$ is present in simulations where the SMBH settled, $\vk<1020\,\kmps$, with the exponent $\nu$ decreasing with increasing $\vk$, in agreement with previous work \citep[e.g.][]{nasim2021b,khonji2024}.
All simulations with $\vk<1020\,\kmps$ display a core in projection however, in agreement with previous observational and theoretical work \citep[e.g.][]{ferrarese1994,nakano1999}.

The presence of a core in the three-dimensional stellar density profile for $\vk=2000\,\kmps$ can be understood due to the lack of an SMBH at the centre of the remnant.
The SMBH, which would ordinarily contribute a point-mass potential that acts as an attractor to stellar particles, is not present, and thus angular momentum-conserving rosette orbits are unable to form or be sustained, thus lowering the central stellar density.
Instead, we observe $\pi$-box orbits to dominate at all radii for the $\vk=2000\,\kmps$ simulation, in tension with the understanding of box orbits being the dominant orbit family to interact with the SMBH binary prior to coalescence. 
However, we find that the excess of $\pi$-box orbits in the $\vk=2000\,\kmps$ simulation primarily originate from angular momentum-conserving orbits (63 per cent) which have reacted to a sudden change in the potential due to the escaping SMBH.

We caution that the claims made in \citet{nasim2021b}, that cores produced by gravitational recoil of the coalesced SMBH are present in the three-dimensional stellar density profile, has the caveat that the recoil velocity must remove the SMBH entirely from the centre of the galaxy remnant, where no oscillations passing through the core are possible.
We find instead that arbitrary recoil kicks, which have a distribution generally peaking well below the escape velocity of the centre of the merger remnant, do not necessarily result in a truly flat three-dimensional stellar density profile, however do produce a flatter density profile compared to the density profile produced by SMBH binary scouring alone.

\subsection{Determining SMBH recoil velocity in observations}
In \autoref{ssec:predcore}, we determined the distribution of possible core radii $\rb$ for our merger remnant by Monte Carlo sampling different SMBH spin vectors prior to SMBH binary coalescence, and using the \citet{zlochower2015} relations to determine the expected recoil velocity.
Our results indicate that GW recoil typically enhances the core region by a factor of 2-3 compared to the core formed through SMBH binary scouring.
However, this information alone cannot be directly used to determine if GW recoil core formation has occurred in a randomly selected galaxy in the real universe, due to the reference radius, $\rbbin$, being in general unknown.

If the core radius of a randomly selected galaxy is significantly enhanced compared to core radii for other galaxies with a similar stellar or SMBH mass, then this galaxy would be an exciting target for assessing the history of GW recoil. 
In \autoref{fig:core-obs} we plot, as a function of the remnant SMBH mass, the median core radii for each of our merger remnants and compare to the observational data presented in \citet{thomas2016}.
The inferred median core sizes range from $\rb\sim0.5\,\mathrm{kpc}$ for the $\vk=0\,\kmps$ simulation, to $\rb\sim1.4\,\mathrm{kpc}$ for the $\vk=1020\,\kmps$ simulation.
Comparing, in \autoref{fig:core-obs}, to the catalogue of core radii in elliptical galaxies presented in \citet{thomas2016} as a function of SMBH mass, we find that the core radii for simulations with $\vk\lesssim240\,\kmps \simeq \sigma_{\star,0}$ are consistent with the observed data to within the intrinsic scatter.
For recoil velocities in excess of $240\,\kmps$, the core radii start to become significantly offset from the data in \citet{thomas2016}.
A similar conclusion is reached when comparing to the core radii data presented in \citet{dullo2014}.

Considering that for our initial conditions, the peak in the transformation-sampled core radius distribution is in the range of $1.0\,\mathrm{kpc}$ - $1.5\,\mathrm{kpc}$ (\autoref{ssec:predcore}, depending on the assumed relation between recoil velocity and core radius), cores in the surface density profile in excess of $1.0\,\mathrm{kpc}$ are viable candidates for GW recoil induced core formation.
Certainly galaxies hosting cores much greater than $2.0\,\mathrm{kpc}$, as for example in the central galaxy A2261-BCG \citep{bonfini2016, dullo2019}, offer prime targets for SMBH recoil kicks to have been a contributing mechanism to the core formation.

Presuming that a galaxy with a large core radius has been found, assessing the magnitude of the SMBH recoil kick can be achieved with the inclusion of IFU observations of the central regions of the galaxy.
As we demonstrated in \autoref{ssec:ifu}, IFU maps uniquely allow for an observational connection to the underlying orbital structure of a galaxy, with imprints of the SMBH binary dynamics apparent in particular in the spatial distribution of $h_4$.
Due to the removal of box orbits from the central regions of the galaxy by the SMBH binary prior to coalescence, a concentration of $h_4\sim0$ appears surrounded by an annulus of $h_4 < 0$.
We propose that identifying the inner radius of the annulus of $h_4<0$ provides a tight estimate on the core radius due to binary scouring, $\rbbin$.

The appearance of $\rbbin$ is particularly apparent in IFU maps restricted to just box orbits. 
Whilst orbital structure cannot be directly observed, but rather inferred through Schwarzschild modelling \citep[e.g.][] {rix1997, cappellari2002, thomas2004} which introduces a new source of uncertainties, generating a model of the observed galaxy to produce mock IFU images with, analogous to the technique presented in this work, would provide a means of determining the radial extent of the annulus of $h_4 < 0$ if it were not apparent in the observed IFU maps.

With an estimate of $\rbbin$, we are able to ascertain the relative size of the observed core radius, $\rb/\rbbin$.
Assuming that a major merger has taken place to produce the observed galaxy, the GW recoil velocity can be inferred by inverting the relation between core radius and recoil velocity, as in our \autoref{eq:vkrb_sig}.
Presumably this relation between $\vk$ and $\rb$ would depend on the mass ratio of the progenitor SMBH binary, so we caution that the relations we provide are only applicable to a major merger involving two equal-mass SMBHs of $M_{\bullet,0}\sim 10^9\,\Msun$.
However equivalent relations may be found through a targeted parameter scan of SMBH recoil velocities, following our methods presented in \autoref{sec:num_sims}.

In the event of a non-monotonic relation between recoil velocity and core radius (unlike our case, where the relation is monotonic), distinguishing the most likely $\vk$ predicted by a single value of $\rb$ may be achieved through assessing the fraction of rosette orbits present in the Schwarzschild modelling of the observed galaxy. 
As demonstrated in \autoref{fig:orbits}, a larger $\vk$ results in less rosette orbits at the centre of the galaxy.

In this way, the SMBH recoil velocity can be well estimated with a synergy of photometric observations (to obtain $\rb$), IFU observations (to determine the spatial distribution of $h_4$), and potentially Schwarzschild modelling (to assess relative orbit fractions).

\subsection{Implications for local cored galaxies}
Our work has important implications for the population of local cored elliptical galaxies.
For elliptical galaxies with similar stellar masses to our simulated mergers ($M_\star \sim 3\times10^{11}\,\Msun$), the observed core radii are generally less than $1\,\mathrm{kpc}$ \citep[e.g.][]{thomas2016,dullo2019}, and the central regions of the galaxy display a tangential bias in the anisotropy parameter $\beta$, seen for example in the SINFONI black hole survey \citep{Thomas2014,saglia2016}.
These observations suggest that following SMBH coalescence, the typical SMBH recoil velocity is low: specifically, $\vk$ is generally less than the stellar velocity dispersion of the SMBH binary-scoured core $\sigma_{\star,0}$.
In our work, whilst the SMBH recoil velocity is set \textit{ab initio}, the only physical mechanism by which it could vary is the orientation and magnitude of the SMBH binary spin vectors.
Zero recoil velocity is achieved through symmetry of the SMBH binary system, and small deviations from symmetry (whether in mass or spin orientations) give rise to small recoil velocities.
Consequently, our work suggests a bias towards near-aligned SMBH spin vectors, rather than SMBH spin vectors orientated randomly on the sphere.
The spin vectors of an SMBH binary can evolve as a result of gas accretion and general relativistic effects.

In the case of gas rich mergers, \citet{miller2013} showed that the spin vectors of the SMBH binary are expected to align when gas accretion onto the SMBHs is prograde, and counteralign when the gas accretion is retrograde.
Similar results have also been found recently by \citet{steinle2023} and \citet{bourne2024}.
In the case of gas-poor mergers, SMBH binary spins likely remain isotropic until entering the GW-dominated hardening phase \citep[e.g.][]{spadaro2024}, with alignment later occurring through general relativistic effects, particularly in the case of unequal-mass SMBHs where the more massive SMBH spin is closely aligned with the orbital angular momentum of the binary \citep{berti2012}.
However, \citet{sayeb2021} found that whilst general relativistic effects can cause precession of the SMBH binary spin vectors, the overall distribution of resulting recoil velocities remains relatively unaffected, owing to the dominant mechanism for SMBH binary spin alignment being the Bardeen-Peterson effect.
The alignment of SMBH binary spin vectors remains an open challenge in modern astrophysics, both observationally and theoretically.

\subsection{Modelling caveats and future work}
It is worthwhile to consider the model limitations of our work, and their potential implications on our conclusions.
As a first limitation, our models do not incorporate gas physics. Work by \citet{blecha2008} and \citet{blecha2011} have suggested that higher gas fractions in merger remnants can act to bring recoiling SMBHs to an equilibrium state more rapidly than in the collisionless case. 
For instance, following the $1:1$ merger of two disc galaxies (leading to a total remnant baryonic mass of $\approx 2\times10^{11}\,\Msun$), \citet{blecha2011} find that for a fixed $v_{\rm kick}$, increasing progenitor disk gas fractions from $0.1$ to $0.3$ can decrease the maximum radius reached by the recoiling SMBH by a factor of $2$-$3$. 
The inclusion of gas in these systems leads to a significant steepening of the central stellar density profile in the remnant due to merger-driven central star formation, as well as additional dynamical friction from the gas itself. 
While \citet{blecha2011} consider systems with similar stellar mass to those studied here, disc-like progenitors are used as opposed to the elliptical progenitors in this work, which for a given stellar mass typically exhibit much lower gas fractions than their disc-like counterparts (e.g. \citealt{Cook2019}). 
\citet{liao2024b} directly show that the differences in central stellar density induced by the inclusion of gas in elliptical-elliptical mergers is negligible compared to the disc-disc case, where the authors specifically match the properties of the disc and elliptical progenitors to observed galaxies. 
As such, for the systems considered in this work, we do not expect gas to play a dominant role in the decay of the kicked SMBH orbit.

The inclusion of hydrodynamics may also serve to alter the stellar environment \textit{prior} to SMBH binary coalescence, affecting when precisely the SMBH binary coalesces.
Higher central stellar densities generally lead to more rapid coalescence of the SMBH binary through three-body interactions and a constantly-refilled loss cone \citep[e.g.][]{chapon2013,souzalima2017,liao2024a}.
We thus have two timescales to consider: the timescale for the galaxy merger remnant to reach an equilibrium state, and the timescale for the SMBH binary to coalesce.
Most importantly, due to uncertainty in the SMBH binary merger timescale driven by eccentricity \citep{nasim2020,rawlings2023} and stellar density \citep{liao2024b}, the triaxiality of the merger remnant at the time of SMBH coalescence is not known a priori.
Additionally, AGN feedback can also alter the central triaxiality of the galaxy merger remnant, though this is a minor contribution in elliptical galaxies \citep{liao2024b}.
A varying triaxiality may impact the asymmetric torquing of the recoiling SMBH, affecting the frequency and timing of the SMBH passages through the stellar core.
Consequently, this might in principle alter the size of the core radius of the merger remnant once the SMBH has settled.
However, as demonstrated in \autoref{fig:trajectory} and discussed in \autoref{ssec:movement}, the passages of the SMBH through the stellar core occur when the SMBH trajectory is predominantly radial, despite a low degree of triaxiality being present in the central regions of the merger remnant at the time of coalescence.
Thus, we do not expect the number of oscillations required for the kicked SMBH to settle to depend strongly on the triaxiality, however the time required for the SMBH to settle may show some variation.
From the analysis of the Lagrangian radii, the later stages of the stellar core growth occur during the passages of the SMBH that take it through the core, and not the total time for the SMBH to settle, thus we expect our results to be robust against triaxiality variation.

In this work, we have only explored one set of initial conditions, and not considered differing mass ratios of progenitor galaxies or SMBHs, nor different stellar density profiles.
Additionally, we consider only a single generation of galaxy mergers, whereas in the cosmological context multiple generations of mergers are expected to have occurred, particularly for massive systems such as the ones we discuss \citep[e.g.][]{Bezanson2009, Naab2009, feldmann2010, vandokkum2015, rantala2024}.
Whilst exploring the full parameter space is beyond the scope of this work, we may consider the results of \citet{khonji2024}, who found that there was no clear trend between SMBH mass and core size from GW recoil, provided the SMBH settled following its initial ejection.
From their work, we can infer that for recoil velocities above $\vk \gtrsim 0.5v_\mathrm{esc}$, the choice of initial conditions does not significantly affect the final core size: this is in contrast to the core size due to binary scouring, which increases with increasing SMBH mass.
The sensitivity of the core size to initial conditions for recoil velocities $\vk \lesssim 0.5v_\mathrm{esc}$ remains to be explored.
One could, for example, use the methodology developed here and apply it to a statistically significant sample of galaxy mergers from cosmological simulations, to investigate the choice of merger initial conditions.
In particular, a statistically significant sample would require both a large number of galaxy mergers, and a representative, diverse sample of galaxy mergers with an occurrence frequency consistent with cosmology.
Such an investigation would not only help address the first limitation raised above, but also answer a new question: what is the overall distribution of core radii in the Universe?

Finally, we make use of the \citet{zlochower2015} relations to determine the SMBH recoil velocity from the pre-merger SMBH binary configuration, which are derived from numerical relativity simulations.
As part of the modelling in this work, we assume that the distribution of spin magnitude $\alpha_\bullet$ at the separation of a binary merger in \ketju{} (some $12\,R_\mathrm{s}$, where $R_\mathrm{s}$ is the Schwarzschild radius of the combined SMBH binary mass) is consistent with the distribution arrived at from numerical relativity, which is necessarily resolved to much smaller scales than our simulations.
Whilst using a different relation than that presented in \citet{zlochower2015} would not affect the kinematic analysis nor the derived relation between recoil velocity and core radius, the distribution of core radii (\autoref{fig:rb_pdf}) would change. 
A distribution of core radii derived from the method proposed in the third limitation discussed would also be sensitive to this.

\section{Conclusions}\label{sec:conclusions}
In conclusion, we find that the GW-induced recoil of an SMBH in a massive elliptical galaxy leaves distinct signatures that would be observable in the real Universe.
Specifically, we find:
\begin{enumerate}
    \item Asymmetric torques on the recoiling SMBH can alter the SMBH trajectory, potentially delaying stellar core growth by repeated passages of the settling SMBH until the SMBH passes through the core.
    \item A core is not present in the three-dimensional stellar density profile unless the SMBH has been ejected entirely from the galaxy, corresponding to the case of zero oscillations of the kicked SMBH about the galaxy centre.
    \item An higher recoil velocity induces a larger core in the projected stellar mass density of the merger remnant than lower recoil velocities, with our models indicating that core radii exceeding $1\,\mathrm{kpc}$ are not only possible, but likely.
    As a consequence, the mass deficit due to SMBH recoil increases with larger recoil velocity, up to $\sim90$ per cent of the coalesced SMBH mass.
    \item For merger remnants with $\vk \lesssim \sigma_{\star,0}$, the majority of stellar mass within the core has angular momentum-conserving orbits. 
    For recoil velocities above this threshold, box orbits which do not conserve angular momentum dominate the mass budget.
    \item Similarly, merger remnants with $\vk \gtrsim \sigma_{\star,0}$ are ergodic within the core, with $\beta\sim 0$.
    Only for the lowest kick velocities are tangentially-biased cores observed, suggesting that the majority of local cored galaxies do not have an SMBH which has experienced a large recoil velocity.
    \item In kinematic observations of the inner region of the merger remnant, the core due solely to SMBH binary scouring, without a contribution from SMBH recoil, can be estimated by identifying the inner radius when the LOSVD symmetric deviation parameter $h_4$ decreases to $h_4 < 0$.
    This in principle allows for an estimate of $\vk$ from the observed core radius.
\end{enumerate}

Taken together, this work establishes a causal link between forward-modelled observations of a galaxy merger remnant -- both photometric and kinematic -- with the complex physics underlying SMBH binary mergers, and their associated GW-induced recoil kicks.
In the future, this will allow us to quantify the recoil velocity of SMBHs in observations of major mergers, opening the way for tighter constraints to be placed on the SMBH binary merger process, particularly for those galaxies in the local Universe.

\section*{acknowledgments}
We thank the anonymous referee for their comments which helped improve the manuscript.
We also thank Jens Thomas for useful discussions on IFU observations and Aki Vehtari for helpful discussions on Bayesian modelling techniques.
A.R. acknowledges the support by the University of Helsinki Research Foundation.
A.R., A.K., M.M., R.J.W., and P.H.J. acknowledge the support by the European Research Council via ERC Consolidator Grant KETJU (no. 818930).
M.M. and P.H.J acknowledge the support of the Academy of Finland grant 339127.
R.J.W. acknowledges the support of the Forrest Research Foundation.
N.K. acknowledges support of the Research Council of Finland Flagship programme: Finnish Center for Artificial Intelligence and the Finnish Foundation for Technology Promotion.
S.L. acknowledges the supports by the National Natural Science Foundation of China (NSFC) grant (no. 12473015, 11988101) and the K. C. Wong Education Foundation.
T.N. acknowledges support from the Deutsche Forschungsgemeinschaft (DFG, German Research Foundation) under Germany’s Excellence Strategy - EXC-2094 - 390783311 from the DFG Cluster of Excellence ``ORIGINS''.

The numerical simulations used computational resources provided by
the CSC -- IT centre for Science, Finland.

\section*{Author contributions}
We list here the roles and contributions of the authors according to the Contributor Roles Taxonomy (\href{https://credit.niso.org}{CRediT}). 
\textbf{AR}: conceptualisation, methodology, formal analysis, investigation, data curation, writing: original.
\textbf{AK}: formal analysis, writing: original.
\textbf{MM}: formal analysis, writing: original.
\textbf{SS}: formal analysis, writing: original.
\textbf{RJW}: investigation, writing: original.
\textbf{NK}: methodology, writing: original.
\textbf{SL}: conceptualisation, writing: review.
\textbf{AR}: writing: review.
\textbf{PHJ}: supervision, writing: original, funding acquisition.
\textbf{TN}: conceptualisation, writing: review.
\textbf{DI}: writing: review.

\section*{Software}
\ketju{} \citep{mannerkoski2023,rantala2017},
\gadget{} \citep{springel2021},
NumPy \citep{harris2020},
SciPy \citep{virtanen2020},
Matplotlib \citep{hunter2007},
pygad \citep{rottgers2020},
\textsc{Stan} \citep{standevelopmentteam2018},
CmdStanPy \citep{standevelopmentteam2018},
Arviz \citep{kumar2019},
Seaborn \citep{waskom2021}.

\section*{Data Availability}
The data underlying this article will be shared on reasonable request to the corresponding author.



\bibliographystyle{mnras}
\bibliography{ref} 




\appendix
\section{Triaxiality}\label{sec:app_triax}
To assess the impact of our (arbitrarily) chosen direction in which the SMBH recoil velocity is applied, we calculate the triaxiality of all merger remnants at the time when the SMBH has settled.

We take all stellar particles within a fraction of the virial radius $r_{200}$, defined as the radius at which the mean density is equal to 200 times greater than the critical density; the virial radius is $\sim450\,\mathrm{kpc}$ for the merger remnants in this study.

\begin{figure}
    \centering
    \includegraphics[width=0.48\textwidth]{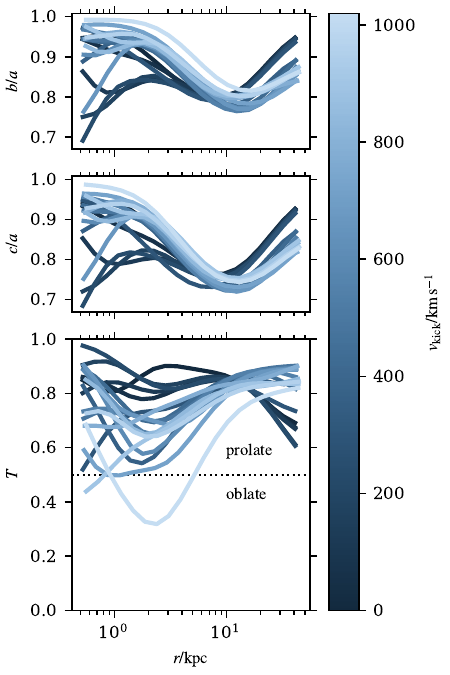}
    \caption{
        Triaxiality ratios $b/a$ and $c/a$ as a function of radius for all merger remnants in the analysis in the top panel and middle panel, respectively.
        Lines are colour coded by the kick velocity, with the same scheme as in \autoref{fig:density3d}.
        Note that the $\vk=0\,\kmps$ case is equivalent to the pre-merger remnant.
        In the bottom panel, the variation of the triaxiality parameter $T$ is shown as a function of radius: the merger remnants in this study are predominantly prolate systems.
    }
    \label{fig:triax}
\end{figure}

We determine the ratios $b/a$ and $c/a$, where $a$, $b$, and $c$ are the eigenvalues of the reduced inertia tensor \citep{gerhard1983,bailin2005}:
\begin{equation}\label{eq:inertia}
    \tilde{I}_\mathrm{red} = \sum_k \frac{r_{k,i} r_{k,j}}{r_k^2},
\end{equation}
and $c\leq b\leq a$. 
The tensor $\tilde{I}_\mathrm{red}$ is determined by binning the stellar component of the remnant into 20 radial shells from $10^{-3} r_{200}$ to $10^{-1} r_{200}$: the binning of particles is not cumulative. 
We apply a uniform filter to reduce Poisson noise due to low particle numbers at small radii.
The radial variations of the ratios $b/a$ and $c/a$ are plotted in \autoref{fig:triax} in the top and middle panels, respectively.
The merger remnants display mildly triaxial cores, albeit without dependence on the recoiling SMBH velocity. 
At radii greater than $4\,\mathrm{kpc}$, the triaxiality of the merger remnants is consistent.
Importantly, the $\vk=0,\kmps$ case also displays only minor variation in $b/a$ and $c/a$ with radius: as this is a good proxy for the pre-merger galaxy remnant, we do not expect triaxiality to dominate over the recoil velocity in the motion of the settling SMBH.

To quantify the nature of the merger remnant triaxiality as a function of radius, we also compute the triaxiality parameter $T$:
\begin{equation}
    T = \frac{1 - (b/a)^2}{1 - (c/a)^2}
\end{equation}
following \citet{franx1991}.
Values of $T$ approaching 1 indicate a prolate system, and values of $T$ approaching 0 indicate an oblate system.
As seen in the bottom panel of \autoref{fig:triax}, at radii $r\gtrsim 5\,\mathrm{kpc}$ all merger remnants are prolate systems. For radii $r\lesssim 5\,\mathrm{kpc}$, only the $\vk=1020\,\kmps$ merger remnant tends to an oblate system.

\section{Core-S\'ersic fit parameter estimates}\label{sec:app_fit}

\begin{figure}
    \centering
    \includegraphics[width=0.48\textwidth]{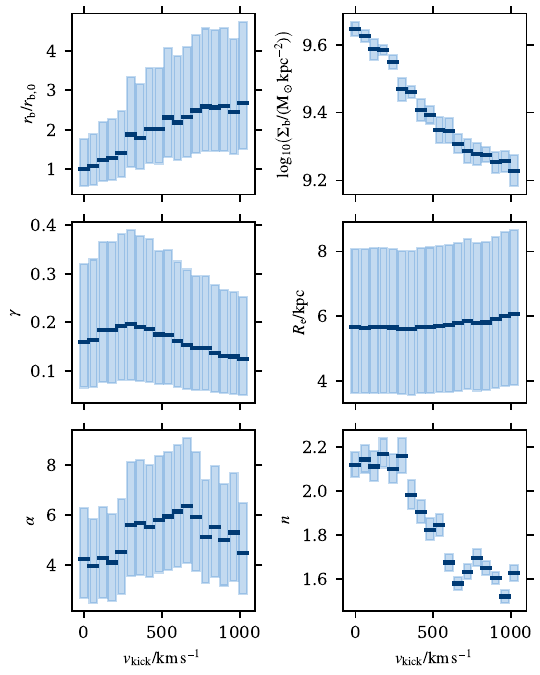}
    \caption{
        Box plots of the six parameters in the core-S\'ersic model (from top left to bottom right) as a function of the kick velocity: the core radius normalised to the pre-merger core radius $\rb/\rbbin$, the density at the core radius $\log_{10}\Sigma_\mathrm{b}$, the core slope $\gamma$, the effective radius $\Reff$, the profile transition index $\alpha$, and the S\'ersic index $n$.
    }
    \label{fig:csparams}
\end{figure}

We use a Bayesian hierarchical model to determine the posterior distributions for the core-S\'ersic projected mass density profile \autoref{eq:cs}.
Bayesian inference, in its simplest form, aims to infer the distribution of the parameter $\theta$ in some model that describe data $y$: this is done with Bayes' formula:
\begin{equation}\label{eq:bayes}
    p(\theta | y) \propto p(\theta) \mathcal{L}(y | \theta),
\end{equation}
where $p(\theta | y)$ is the posterior distribution, $p(\theta)$ is the prior distribution that encodes belief about how the parameter $\theta$ is distributed, and $\mathcal{L}(y | \theta)$ is the likelihood of observing the data $y$ given the model parameter $\theta$.
In the case that there are groups of data that are subsets of some global population, and these subsets are exchangeable in the sense that there is no distinguishing order to them, we can model each subset of data $y_j$ as having its own value of the model parameter $\theta_j$.
The distribution of all values of $\theta_j$ can then be modelled as samples from a global distribution, termed a hyperdistribution, with hyperparameter(s) $\phi$: this is where the hierarchical nature of the model is introduced.
A prior distribution can then be assigned to the hyperparameter $\phi$, $p(\phi)$.
As such, \autoref{eq:bayes} then takes the form:
\begin{equation}
    p(\theta, \phi | y) \propto p(\phi) p(\theta|\phi) \mathcal{L}(y | \theta, \phi).
\end{equation}

In our application of inferring the parameter vector $\vb{\theta}\equiv[\rb, \Sigma_\mathrm{b}, \gamma, \Reff, n, \alpha]$ that describes the projected stellar mass density profile of the galaxy merger remnant, each projection from which the merger remnant is viewed from constitutes a subset of the surface density data $\Sigma_j$.
As each of the projections are unordered and exchangeable (i.e., there is no distinguishing label for the projections), and each projection is of the same merger remnant, the hierarchical model naturally lends itself.
For each projection, we thus infer the six parameters of $\vb{\theta}$, and assume that these values are draws from a common, global distribution of all possible $\vb{\theta}$ vectors.
Hence, we are effectively inferring the global hyperparameters $\vb{\theta}^\mathrm{hyp}$ that describe the latent parameters $\vb{\theta}$ that we are interested in.
This idea is demonstrated in the directed acyclic graph (DAG) for the model, shown in \autoref{fig:dag}.
The posterior density can thus be constructed for each projection- and radially-dependent surface density element $\Sigma_{i,j}^\mathrm{L} \equiv \log_{10}(\Sigma_{i,j})$ as:
\begin{multline}
    p(\vb{\theta}, \vb{\theta}^\mathrm{hyp} | \Sigma_{i,j}') \propto p\left(\vb{\theta}^\mathrm{hyp}\right) p\left(\vb{\theta} | \vb{\theta}^\mathrm{hyp}\right) \\ \times \prod_j^{N_\mathrm{proj}} \prod_i^{N_R} \mathcal{N}\left(\Sigma_{i,j}^\mathrm{L} | \hat{\Sigma}^\mathrm{L}\left(R_i | \vb{\theta}_j, \vb{\theta}^\mathrm{hyp}\right), \tau_i\right),
\end{multline}
where we use a normal distribution\footnote{Note that the normal distributions $\mathcal{N}$ are parameterised using the standard deviation $\sigma$ in this work, as opposed to the variance $\sigma^2$ as is common in many statistics references.} as the likelihood function, and the error in the fit depends on the radial bin $i$.

We use weakly-constrained prior distributions for each of the hyperparameters in $\vb{\theta}^\mathrm{hyp}$, where these prior distributions are chosen to reflect reasonable values \citep{gabry2019} that the parameters may take (e.g. distributions of a radial quantity are chosen so as to be positively-constrained).
The distributions of the hyperparameters are given in \autoref{tab:hyper}.

We fit the model with \textsc{Stan} \citep{standevelopmentteam2018} using the No U-turn sampler (NUTS, a variant of Hamiltonian Monte Carlo (HMC)) with four chains each of 4000 iterations, of which the first 2000 are discarded as warmup.
This resulted in 8000 draws from the posterior distribution of $\vb{\theta}$, whereby we ensure the quality of the HMC fit by enforcing a diagnostic value $\hat{R}$ (comparison of between-chain and within-chain variance) less than 1.05, the effective sample size is high, and that the number of divergent transitions is less than 5 per cent \citep[for details see][]{vehtari2021,standevelopmentteam2018}.
We perform additional prior sensitivity tests using the power-scaling method described in \citet{kallioinen2024}, ensuring that the sensitivity diagnostic for prior power-scaling is less than 0.05 for the latent parameters $\vb{\theta}$ of the model. 
This recommended threshold indicates that the sampled posterior distributions of the latent parameters are not sensitive to the particular prior distributions used.

In \autoref{fig:csparams}, we show as box plots the distributions of the six latent parameters in the core-S\'ersic profile, \autoref{eq:cs}.
The dark bars indicate the median of each distribution, and the boxes are bounded at the bottom by the $\ordinal{25}$ percentile, and capped at the $\ordinal{75}$ percentile.
As discussed in the main text, the median core radius $\rb$ (top left) increases in both magnitude and variance with increasing kick velocity.
The density at the core radius (top left) decreases with increasing kick velocity, as well as having a variance which is largely independent of the kick velocity.
The slope of the inner density profile, $\gamma$ (centre left) is largely independent of the kick velocity, as is the effective radius of the galaxy (centre right) and transition index (bottom left).
The S\'ersic index (bottom right) is constant for $\vk \leq 300\,\kmps$, and is constant, albeit with a smaller value, for $\vk \geq 660\,\kmps$.
For the intermediate kick velocities, there appears to be a linear decrease in the S\'ersic index $n$.

\begin{figure*}
    \centering
    \includegraphics[width=\textwidth]{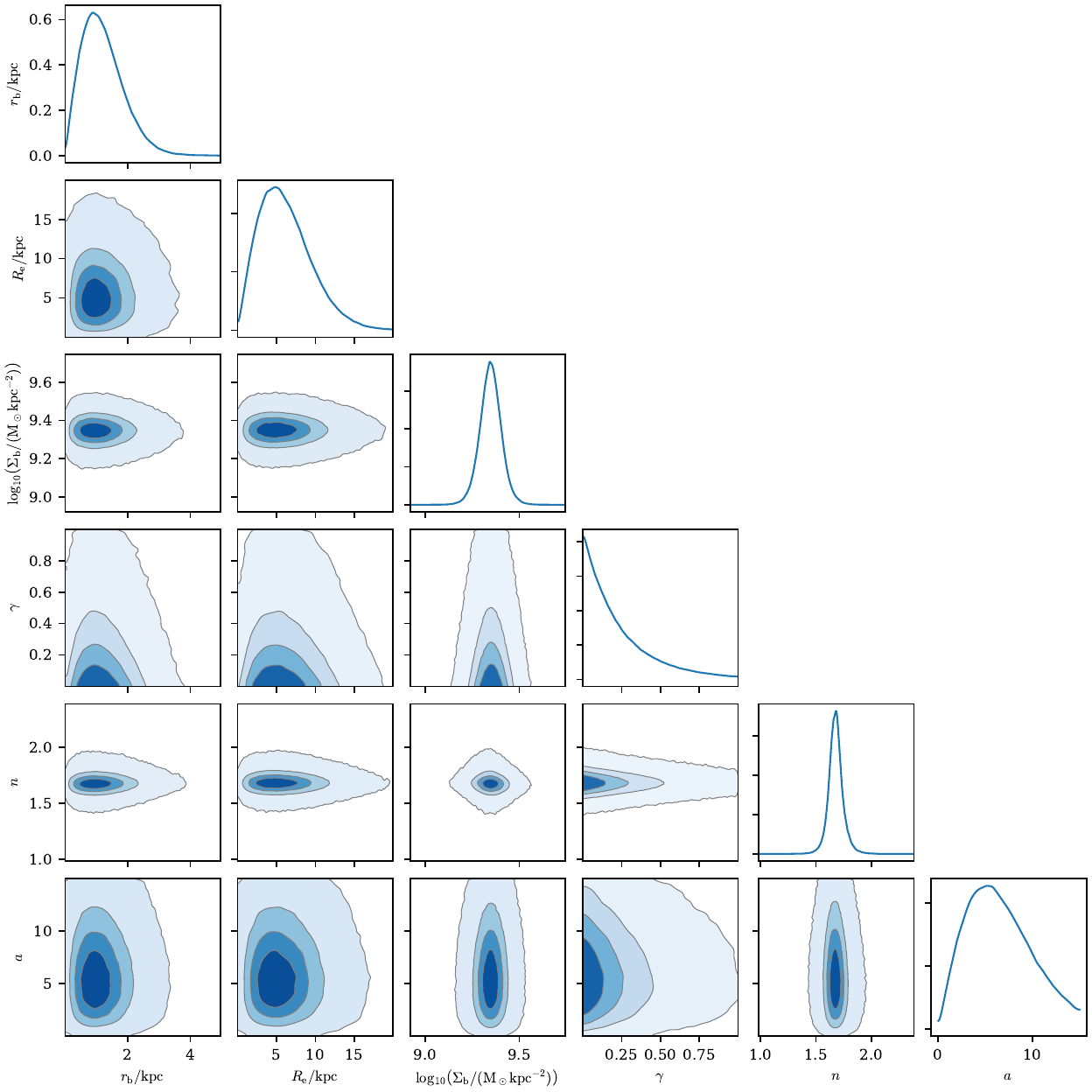}
    \caption{
        Corner plot of the latent parameters in the core-S\'ersic model for a representative example, here the $\vk=600\,\kmps$ instance.
        Contours indicate, from dark to light blue, the 25 per cent, 50 per cent, 75 per cent, and 99 per cent highest density intervals (HDIs).
        All distributions are unimodal with little cross-correlation between the latent variables.
    }
    \label{fig:cscorner}
\end{figure*}

The joint relation between the latent parameters in \autoref{eq:cs} can be seen in the representative example of $\vk=600\,\kmps$ in \autoref{fig:cscorner}.
The leading diagonal shows the marginal distributions of each latent parameter; each distribution is unimodal.
The contours in the lower diagonal triangle show the joint posterior distribution between each of the latent parameters in \autoref{eq:cs}, with the contour colouring indicating various highest density intervals (HDIs).
In no pairing of parameters is a strong covariance observed; the marginal distributions are orthogonal to each other.
One case worth a special mention is the width of the 25 per cent HDI for the parameter $\alpha$.
The wide HDI indicates that a large range of $\alpha$ values maximise the posterior distribution.
However, as discussed in \citet{graham2003}, values of $\alpha\gtrsim 3$ have a similar effect to induce a sharp transition between the inner and outer surface density profiles.
The variation of $\alpha$, particularly to large values greater than 10, are not expected to have a strong effect on the overall surface density profile.

\section{Supporting information for IFU analysis}\label{sec:app_ifu}
\begin{figure}
    \centering
    \includegraphics[width=0.5\textwidth]{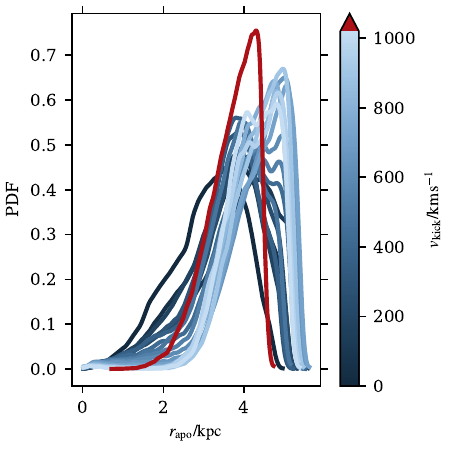}
    \caption{
        The distribution of apocentres for stellar orbits with a mean position within $0.25r_{1/2}$ (matching the mock IFU maps), determined using the orbit analysis routine described in \autoref{sec:orbits}.
        All the distributions are left-skewed, where simulations with higher $\vk$ are more left-skewed than those with lower $\vk$.
    }
    \label{fig:apo_dist}
\end{figure}

In \autoref{fig:apo_dist}, we show, for each recoil velocity,  kernel density estimates of the apocentre distributions for stellar particles within a mean position of $0.25r_{1/2}$ from the merger remnant centre.
The distributions are clearly left-skewed, and becoming increasingly skewed as the recoil velocity increases, with the exception of the $\vk=2000\,\kmps$ simulation, as discussed in the main text.

\begin{figure}
    \centering
    \includegraphics[width=0.5\textwidth]{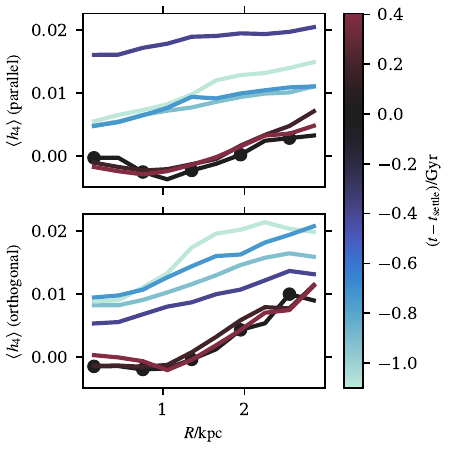}
    \caption{
        The dependence of the $\langle h_4 \rangle$ profile on the time of analysis for a representative example $\vk=900\,\kmps$.
        At early times, when the SMBH is still oscillating through the stellar medium, the $\langle h_4 \rangle$ profile is overall elevated, and decreases as the SMBH begins to settle towards an equilibrium point.
        For times after the SMBH has settled (according to the definition presented in \autoref{ssec:settle}), the profiles are consistent in both the parallel and orthogonal projections of $\langle h_4 \rangle$.
        The profile corresponding to $t-t_\mathrm{settle}=0\,\mathrm{Gyr}$ is accentuated with circle markers.
    }
    \label{fig:h4900}
\end{figure}

In \autoref{fig:h4900}, we show how the radial dependence of $\langle h_4 \rangle$ varies with regard to the time when the SMBH has settled, defined $t-t_\mathrm{settle}=0\,\mathrm{Gyr}$, for a representative example ($\vk=900\,\kmps$).
At the time the SMBH settled, and times following this point, the radial dependence of $\langle h_4 \rangle$ is time-invariant, for both the parallel and orthogonal projections.
This ensures that our conclusions in the main text are not sensitive to the exact moment the LOSVD is constructed.

\bsp	
\label{lastpage}

\end{document}